\documentclass{pas}
\usepackage{microtype}
\usepackage{booktabs}
\usepackage{xcolor}
\usepackage{natbib}

\newcommand{\aap}{A\&A}
\newcommand{\apjl}{ApJ}

\newcommand{\apjs}{ApJS}

\usepackage{multirow}
\usepackage{graphicx}
\usepackage{appendix}
\usepackage{etex}
\usepackage{gensymb}
\usepackage{amsmath,siunitx}
\usepackage [english]{babel}
\usepackage [autostyle, english = american]{csquotes}
\MakeOuterQuote{"}

\begin{document}

\lefttitle{}
\righttitle{Ezhikode et al.}

\jnlPage{1}{11}
\jnlDoiYr{2024}
\doival{xx.xxxx/pasa.xxxx.xx}

\articletitt{Research Paper}

\title{Serendipitous detection of an intense X-ray flare in the weak-line T~Tauri star KM~Ori with SRG/eROSITA}

\author{\sn{Savithri H.} \gn{Ezhikode}$^{1,2}$, \sn{Hema} \gn{Anilkumar}$^{1}$, \sn{R.} \gn{Arun}$^{3}$, \sn{Blesson} \gn{Mathew}$^{1}$, \sn{V.} \gn{Jithesh}$^{1}$, \sn{Suman} \gn{Bhattacharyya}$^{1}$, \sn{Sneha} \gn{Nedhath}$^{3,1}$, \sn{T. B.} \gn{Cysil}$^{1}$, \sn{S.} \gn{Muneer}$^{3}$, \sn{Sreeja S.} \gn{Kartha}$^{1}$, and \sn{Pramod} \gn{Kumar S}$^{3}$ }

\affil{$^1$Department of Physics and Electronics, CHRIST (Deemed to be University), Bangalore 560029, India\\$^2$St. Francis de Sales College (Autonomous), Electronics City, Bengaluru 560100, India\\$^3$Indian Institute of Astrophysics, Koramangala, Block II, Bangalore 560034, India}

\corresp{Savithri H. Ezhikode, Blesson Mathew, Email: savithrih@sfscollege.in, blesson.mathew@christuniversity.in}

\begin{abstract}
Weak-line T Tauri stars (WTTS) exhibit X-ray flares, likely resulting from magnetic reconnection that heats the stellar plasma to very high temperatures. These flares are difficult to identify through targeted observations. Here, we report the serendipitous detection of the brightest X-ray flaring state of KM~Ori in the eROSITA DR1 survey. Observations from SRG/eROSITA, \textit{Chandra X-ray Observatory}, and \textit{XMM-Newton} are analysed to assess the X-ray properties of KM~Ori, thereby establishing its flaring state at the eROSITA epoch. The long-term (1999--2020) X-ray light curve generated for the Chandra observations confirmed that eROSITA captured the source at its highest X-ray flaring state recorded to date. Multi-instrument observations support the X-ray flaring state of the source, with time-averaged X-ray luminosity ($L_{\rm 0.2-5~keV}$) reaching $\sim 1.9\times10^{32}\rm{erg~s^{-1}}$ at the eROSITA epoch, marking it the brightest and possibly the longest flare observed to date. Such intense X-ray flares have been detected only in a few WTTS. The X-ray spectral analysis unveils the presence of multiple thermal plasma components at all epochs. The notably high luminosity ($L_{\rm 0.5-8~keV}\sim10^{32}~{\rm erg~s}^{-1}$), energy ($E_{\rm 0.5-8~keV}\sim10^{37}$~erg), and the elevated emission measures of the thermal components in the eROSITA epoch indicate a superflare/megaflare state of KM~Ori. Additionally, the H$\alpha$ line equivalent width of $\sim$-5~\AA~ from our optical spectral analysis, combined with the lack of infrared excess in the spectral energy distribution, were used to re-confirm the WTTS (thin disk/disk-less) classification of the source. The long-duration flare of KM~Ori observed by eROSITA indicates the possibility of a slow-rise top-flat flare. The detection demonstrates the potential of eROSITA to uncover such rare, transient events, thereby providing new insights into the X-ray activity of WTTS.
\end{abstract}


\maketitle
\section{Introduction}
\label{intro}

T Tauri stars (TTS) are young, low-mass ($M_\star \leq 2~M_\odot$) pre-main-sequence stars (PMS), characterised by variability, strong chromospheric activity, and emission lines such as H$\alpha$. They can be classified into Classical T Tauri stars (CTTS) and Weak-line T Tauri stars (WTTS) based on the presence of the circumstellar disk. WTTS lack prominent circumstellar disks \citep{1993AJ....106.2005G}, making them crucial for studying early stellar evolution without the complications of disk accretion. Unlike CTTS, WTTS do not exhibit obvious signs of accretion or possess optically thick disks, suggesting they represent a later evolutionary stage where the circumstellar disk has either become optically thin or dispersed. Despite this, some WTTS are located in the same region of the Hertzsprung-Russell diagram as CTTS, which indicates that there is no specific preferred disk lifetime \citep{1988AJ.....96..297W,1997A&A...319..184A,2003Ap.....46..506P, 2021AcAT....2a...1P}. WTTS emit X-rays through magnetic reconnection events, which heat the stellar plasma to extremely high temperatures \citep{2004A&A...427..263F, Feeney2021A&A...653A.101F}. X-ray flares in WTTS are characterised by rapid decay and lower detection frequency (e.g., \citealp{Stelzer2000A&A...356..949S}). Thus, detecting them via targeted observations is very challenging. Despite this, investigating WTTS flares remains crucial for understanding the nature of magnetic activity in the later phases of star formation.

X-ray surveys from various missions provide ample opportunities to study X-ray emission from diverse stellar samples. The extended ROentgen Survey with an Imaging Telescope Array \citep[eROSITA;][]{2021A&A...647A...1P}, the primary instrument onboard the \textit{Spektrum-Roentgen-Gamma} \citep[SRG;][]{2021A&A...656A.132S} mission, is critical in this context as it is designed to conduct a highly sensitive all-sky survey in the soft to medium X-ray bands (0.2--10 keV). The wide field of view, repeated sky scans, and spectral capabilities make eROSITA well-suited to discover and study X-ray flares from the WTTS population. In a quest to study the X-ray sources in the Orion Complex, we identified the WTTS KM~Ori in a very high X-ray flux state while cross-matching the APOGEE-2 survey \citep{2018AJ....156...84K} catalogue with the SRG/eROSITA all-sky survey (eRASS1) Data Release 1 \citep[DR1;][]{2024A&A...682A..34M} catalogue.

KM~Ori is a young, K5-type WTTS within the Orion star-forming complex \citep{2019MNRAS.490.3158C}. Its optical spectrum shows weak H$\alpha$ emission, a characteristic of WTTS \citep{1994AJ....108.1906H, 2013ApJS..208...28S}. Spectroscopic studies \citep[e.g.,][]{2017A&A...608A..77L, 2020ApJ...888..116S} reveal a moderate surface magnetic field with a strength of approximately 1.9~kG. KM~Ori exhibits significant X-ray emission \citep{2005ApJS..160..319G, 2008ApJ...677..401P}, characterised by flaring events \citep{2008ApJ...688..418G}. Observations, including those from the \textit{Chandra} Orion Ultradeep Project (COUP), have been pivotal in characterising X-ray variability and luminosity in KM~Ori \citep[e.g.,][]{2005ApJS..160..319G, 2016A&A...587A..81B}. Additionally, the rotational modulation of X-ray emission, as seen in the COUP data \citep{2005ApJS..160..450F}, suggests a spatially inhomogeneous distribution of magnetically active regions on its surface.

In this paper, we report the eROSITA detection of the WTTS KM~Ori in its highest X-ray flaring state ever recorded. We characterise the properties of this observation and compare them with previous observations during flaring and quiescent states, conducted with \textit{Chandra} and \textit{XMM-Newton}, which we discuss in the subsequent sections. The muti-wavelength/mission data used for this work is explained in \S\ref{data}. The X-ray, optical, and SED analyses of the source are detailed in \S\ref{analysis}. A brief discussion, along with major conclusions, is included in the last section (\S\ref{rslt_disc}).

\section{Observations and Data Reduction}
\label{data}

\subsection{X-ray}

This study utilises X-ray data from observations conducted with SRG/eROSITA, \textit{Chandra X-ray Observatory (CXO)}, and \textit{XMM-Newton}. Details of the observations and reduction methods applied to these data sets are given below. In order to identify the presence of any previous flare in the source, we obtained the Chandra observations of the source observed from 1999 to 2020. However, we observed a strong flare only in one of the observations during the COUP survey. We discuss this in detail in \S\ref{sec_lc}. Since our aim is to characterise the X-ray flare in the eROSITA epoch (X1), we compare the relevant \textit{Chandra} observation conducted on 10 January 2003 (X2) and the lowest flux state observation with \textit{Chandra} conducted on 27 November 2019 (X3) for further analysis. Also, we utilise the longest available \textit{XMM-Newton} observation, conducted on 2 March 2006 (X4) to obtain high-quality spectrum. This long observation did not show the presence of any flaring events in the source. Hence, these two data represent previous observations at flaring and quiescent epochs. Table~\ref{tab:obs} collates the details of these observations, including mission information, date of observation and exposure time.

\subsubsection{SRG/eROSITA}

For this study, we used the eROSITA data of KM Ori from the DR1 release observed during the eRASS1 survey. The data release encompasses observations of the Western Galactic hemisphere within the longitude range of $359.94 > l >  179.94$~degrees. The data products of the observation of KM Ori \citep[DR1 main catalogue;][]{2024A&A...682A..34M} were retrieved from the eROSITA-DE DR1 archive (eRODat\footnote{https://erosita.mpe.mpg.de/dr1/erodat/}). The data products comprise the source and background spectra, light curves, and response files for the seven individual Telescope Modules (TM) and the merged TM combination. We used the science data products obtained for the combined telescope modules for the subsequent analyses. The spectrum was grouped to ensure a minimum of 20 counts per bin. The details of spectral and light curve analyses are given in \S\ref{xry}.

\subsubsection{\textit{Chandra}/ACIS}

The ONC region has been extensively investigated \citep{2003ApJ...582..382Flaccomio,2005ApJS..160..319G} using the Advanced Camera for Imaging Spectroscopy (ACIS) onboard the \textit{Chandra}. We retrieved data for all available observations of the field obtained with Chandra to see if KM~Ori was flaring in any of these epochs. There are 87 observations spanning from 1999 to 2020, and KM~Ori was detected in 60 observations. We obtained the level-2 data of these observations from the public \textit{Chandra} Data Archive and reprocessed using the software package CIAO \citep{10.1117/12.671760Fruscione} version 4.16, along with the calibration database (CALDB) 4.11.0. To the reprocessed level-2 data (event file), an energy filter of 0.3 $-$ 8 keV is applied. These event files were then checked for flaring background using \textit{deflare} task. We found background flaring in two observations, which were cleaned to contain only good time events. Certain observations were affected by readout streaks, which hindered the effectiveness of the background flare cleaning process. To resolve this, we utilised the {\sc acis\_streak\_map} tool to create a detailed map of the readout streaks associated with bright sources. This map enabled us to identify and eliminate the streaks, thereby enhancing the reliability of our background analysis. We used the source detection algorithm {\sc wavdetect} \citep{2002ApJS..138..185Freeman} to detect the sources within the CCD. Wavelet scales ranging between 1 and 32 in steps of 2$^{\rm n}$ (where n = 0,1,2,3,4,5) were used. To ensure reliable detections and minimize false positives, a stringent significance threshold of 10$^{-6}$ was set. A circular region with a radius of 6~arcsecs was defined for the source to encompass 90\% of the PSF at 1.5 keV, as determined by the {\it wavdetect} algorithm. A circular region of radius $\sim$~30~arcsecs devoid of sources on the same CCD chip was selected for background measurements.

\subsubsection{\textit{XMM-Newton}/EPIC}

\textit{XMM-Newton} observed the region containing the source for a duration of about 94 ks on March 2, 2006. We retrieved the X-ray data from the European Photon Imaging Camera \citep[EPIC;][]{2001A&A...365L..18S} PN and MOS detectors, which were observed in full window mode with a medium filter. We performed the data reduction using SAS version 21.0, employing the {\it epproc} and {\it emproc} tasks and utilising updated calibration files as of March 2024. We identified and eliminated the intervals of flaring particle background from the event lists based on threshold rates (0.5~counts~s$^{-1}$ for PN, 0.2~counts~s$^{-1}$ for MOS1, and 0.23~counts~s$^{-1}$ for MOS2) derived from high-energy light curves. Event lists were filtered to retain the highest quality events (FLAG=0) with PATTERN$<=$4 (single and double events) for PN and PATTERN<=12 (up to quadruples) for MOS. Spectra were extracted from circular regions with a radius of 24~arcsecs centred on the source, while background spectra were extracted from multiple circular regions (28~arcsecs radii) on the same CCD chip. We generated the spectra and response files (ARF and RMF) for the cleaned event lists using the {\it xmmselect} task. We further re-binned the spectra and linked them to the response and background files using the {\it specgroup} tool. This process ensured a minimum of 25 counts per bin, with an oversampling factor of 5 applied. The background corrected light curves were generated using the {\it epiclccorr} task in SAS.

\begin{table*}[h]
    \centering
    \caption{Details of X-ray and optical observations. The last column shows the net exposure time for each observation in the corresponding energy bands used for the spectral analysis.}
    \begin{tabular}{ccccccc}
    \hline
    \hline
        Mission & Obs Date & OBS ID & OBS & Instrument & Exposure (ks) \\
        \hline
        \hline
        &  &  X-ray  &  &   \\
        \hline
        SRG & 2020-03-21 & 108309600003 & X1 & eROSITA & 0.2 \\
        Chandra & 2003-01-10 & 3744 & X2 & ACIS-I & 83.9 \\
               & 2019-11-27 & 23008 & X3 & ACIS-S & 47.4 \\
        XMM-Newton & 2006-03-02 & 0403200101 & X4 & EPIC (PN) & 52.1  \\
         &  & & & EPIC (MOS1) & 72.7  \\
         &  & & & EPIC (MOS2) & 75.2  \\
        \hline
        
        &  &  Optical  &  &   \\
        \hline
        HCT & 2024-02-27 & & O1 & HFOSC/Gr7 & 0.6  \\
            & 2024-03-25 & & O2 & HFOSC/Gr7 & 0.6 \\
        \hline
        \hline
    \end{tabular}
    \label{tab:obs}
\end{table*}

\subsection{Optical}
We conducted low-resolution optical spectral observations of KM~Ori using the HFOSC instrument on the 2.01 m Himalayan Chandra Telescope (HCT) at two epochs on February 27, 2024, and March 25, 2024. The observations were carried out with grism 7 (Gr7) covering the wavelength range of 3800--6840~{\AA}, employing the 167l slit. Standard IRAF \citep{1986SPIE..627..733T} routines were used for spectral analysis. Spectral data obtained from both epochs exhibited a resolution of R$\sim$1000.

\section{Analysis and Results}
\label{analysis}

\begin{figure}[h]
    \centering
    \includegraphics[scale=0.5]{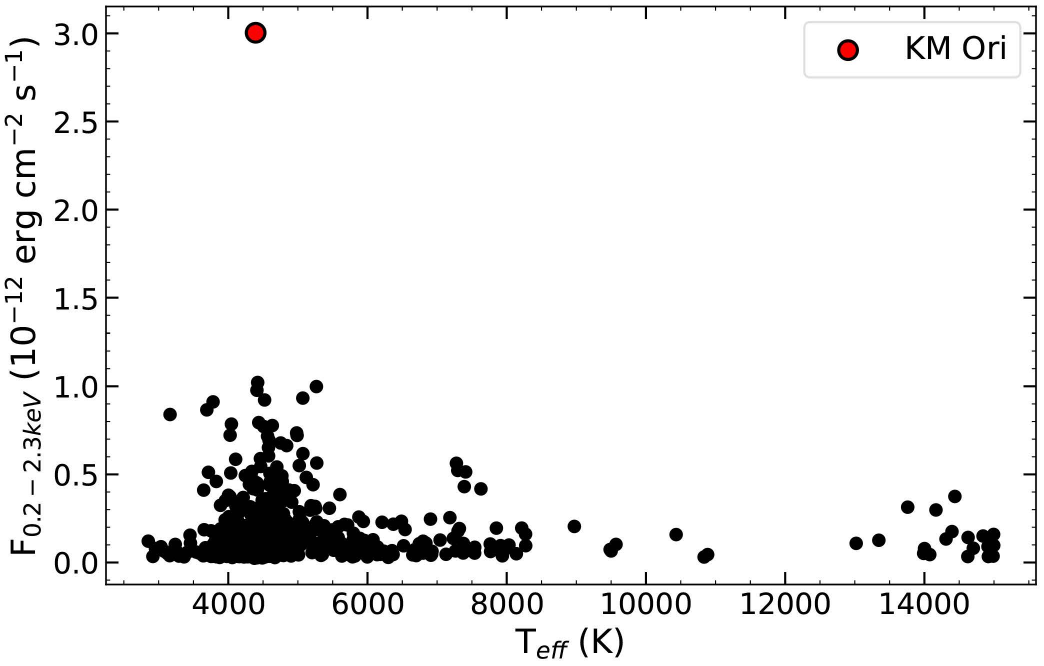}
    \caption{The figure shows the eRASS1 flux (0.2--2.3~keV) versus the stellar effective temperature of APOGEE-2 sources in the Orion Complex.}
    \label{fig:orion}
    \vspace*{-0.5cm}
\end{figure}

Building upon the APOGEE-2 survey of the Orion Complex (Kunkel et al. 2018), we have constructed a catalogue of 1030 sources with both eRASS1-detected X-ray counterparts and APOGEE-derived stellar parameters. The eROSITA flux in the 0.2-2.3~keV band (Fig.~\ref{fig:orion}) reveals KM~Ori as the most luminous X-ray source in our sample, $\sim30$ times the median flux (${\rm 0.097\times10^{-12}~erg~cm^{-2}~s^{-1}}$) for the complex. The presence of a hard X-ray (2.3--5~keV) component in the eRASS1 catalogue (UID\_Hard = 208309600003) identifies KM~Ori as a particularly compelling target, motivating us to conduct an in-depth X-ray and optical investigation, which is presented below.

\subsection{X-ray Analysis}
\label{xry}

\begin{figure}[h]
    \centering
    \includegraphics[scale=0.45]{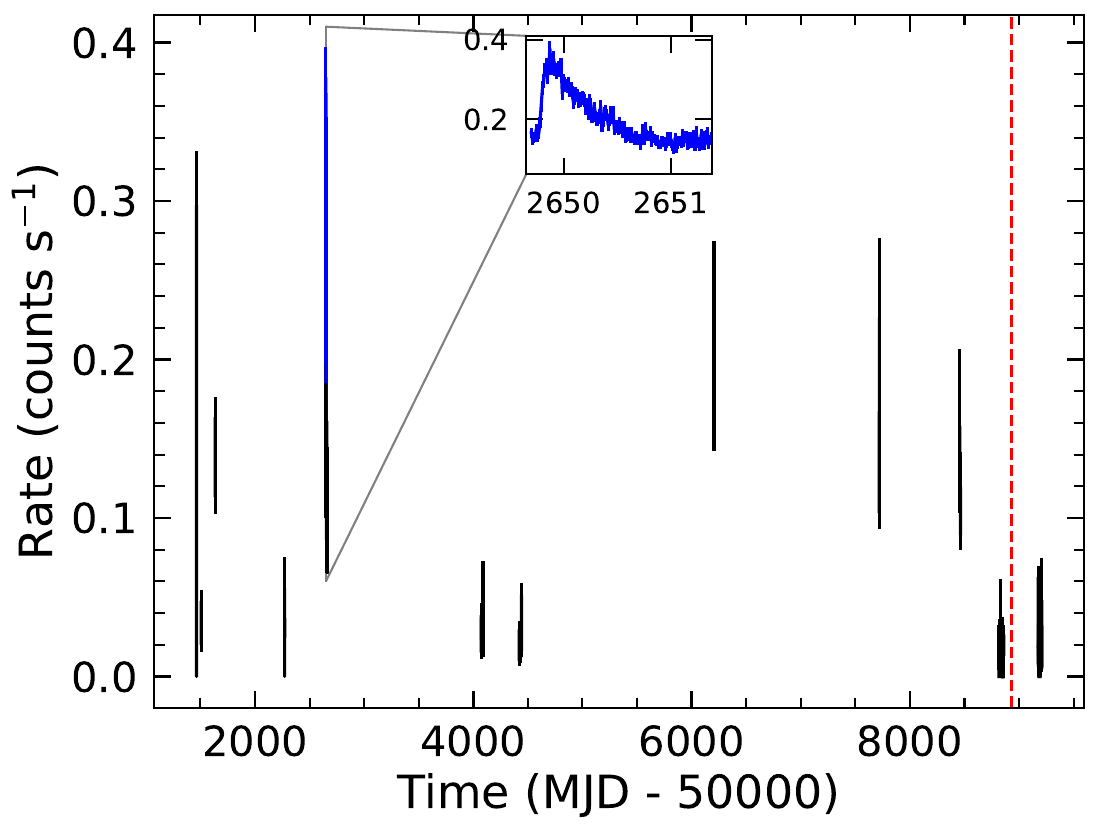}
    \caption{Long-term \textit{Chandra} light curve (0.2--5~keV) of KM~Ori, spanning from 1999-10-12 to 2020-12-25. The inset shows the light curve extracted for the highest X-ray flux state, which is observed in the COUP survey (ObsID: 3744). The red dashed vertical line represents the epoch of eROSITA observation.}
    \label{fig:chandra_lc}
\end{figure}

\subsubsection{Light curves}
\label{sec_lc}

The eRASS1 light curve, included in the standard data products, encompasses time series data across three distinct energy bands (0.2--0.6~keV, 0.6--2.3~keV, and 2.3--5~keV), binned at 10~s intervals. We utilised the \emph{python} code \emph{eRebin} provided by \cite{2021Natur.592..704A, 2024A&A...684A..64A} to create rebinned source-only (background subtracted) light curves for different energy bands provided in the standard light curve. The code rebins the light curves for 40~s, which is the on-target exposure of an \textit{eROday} (4~hours). We added these light curves using {\it lcmath} to generate a total light curve in the 0.2--5~keV band. The gaps of $\sim4$~hours seen in the eROSITA light curve correspond to one revolution of the spacecraft in all-sky survey mode. The light curve showed moderate variability with fractional rms variation $F_{rms}=0.20\pm0.07$ across the $\sim$86~ks observation.

We extracted light curves for all \textit{Chandra} observations of KM~Ori in the 0.2--5~keV band. The light curve spanning from 1999 to 2020 is shown in Fig.~\ref{fig:chandra_lc}. The highest flux state within this long-term dataset was observed during the COUP observation with Obs ID 3744, which is highlighted in the inset of the figure. To compare the eROSITA flare observation with previous flare data from the source, we focused our analysis on this specific Chandra observation (X2). Additionally, we analysed the Chandra observation with the lowest count rate (X3) and the longest available XMM-Newton observation (X4). The Chandra light curve of observation X2 ($F_{rms}=0.32\pm0.02$) clearly displayed flaring activity, consistent with previous reports by \cite{2005ApJS..160..319G, 2008ApJ...688..418G}. The flaring was observed within the initial $\sim$85~ks of the $\sim$165~ks long \textit{Chandra} observation. On the other hand, the light curves for X3 (duration $\sim$ 48~ks) and X4 (duration $\sim$ 94~ks) did not exhibit significant variability ($F_{rms}<0.1$ and 0.13$\pm$0.02, respectively). 

\begin{figure*}[h]
    \centering
    \includegraphics[width=0.65\columnwidth]{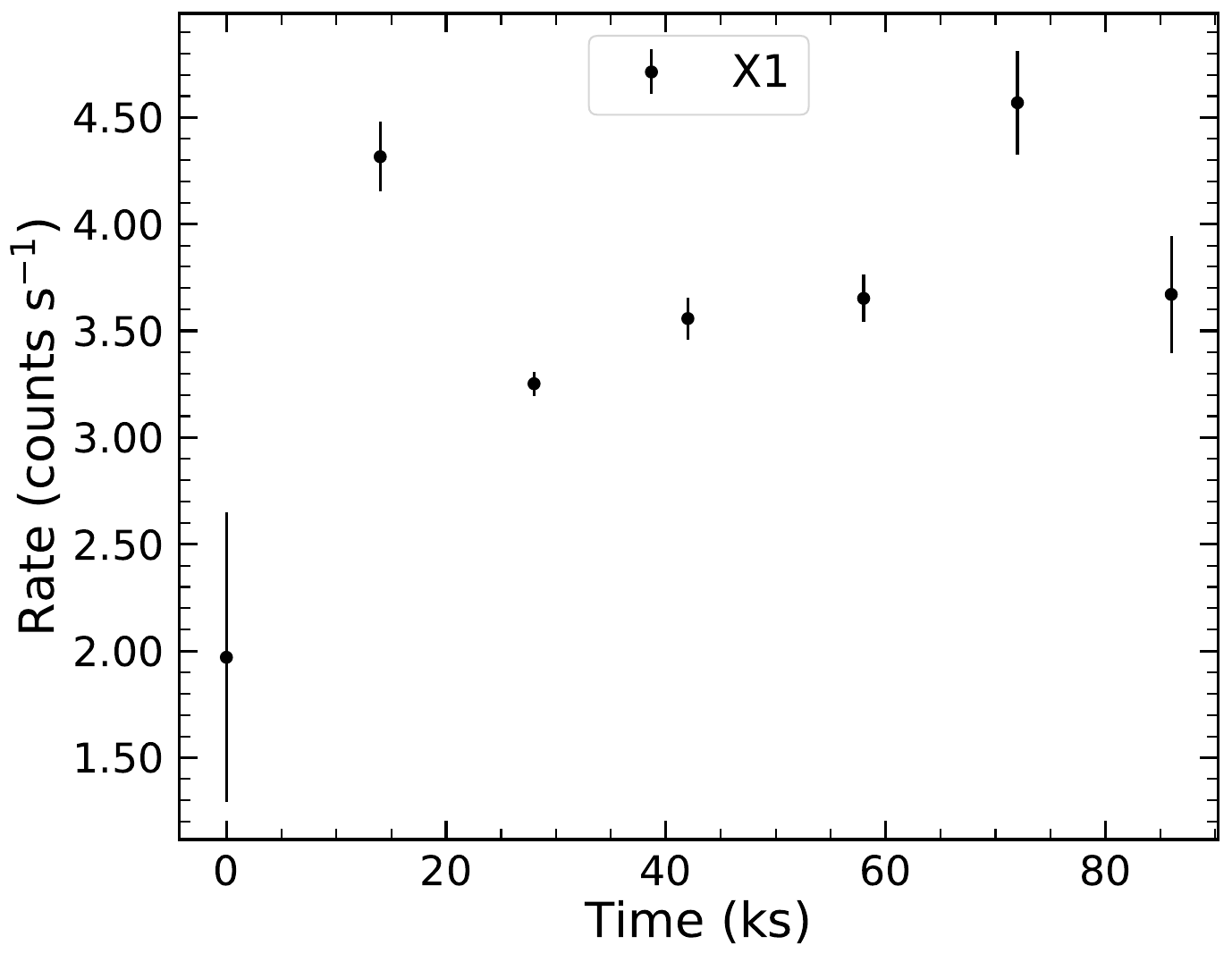}
    \includegraphics[width=0.65\columnwidth]{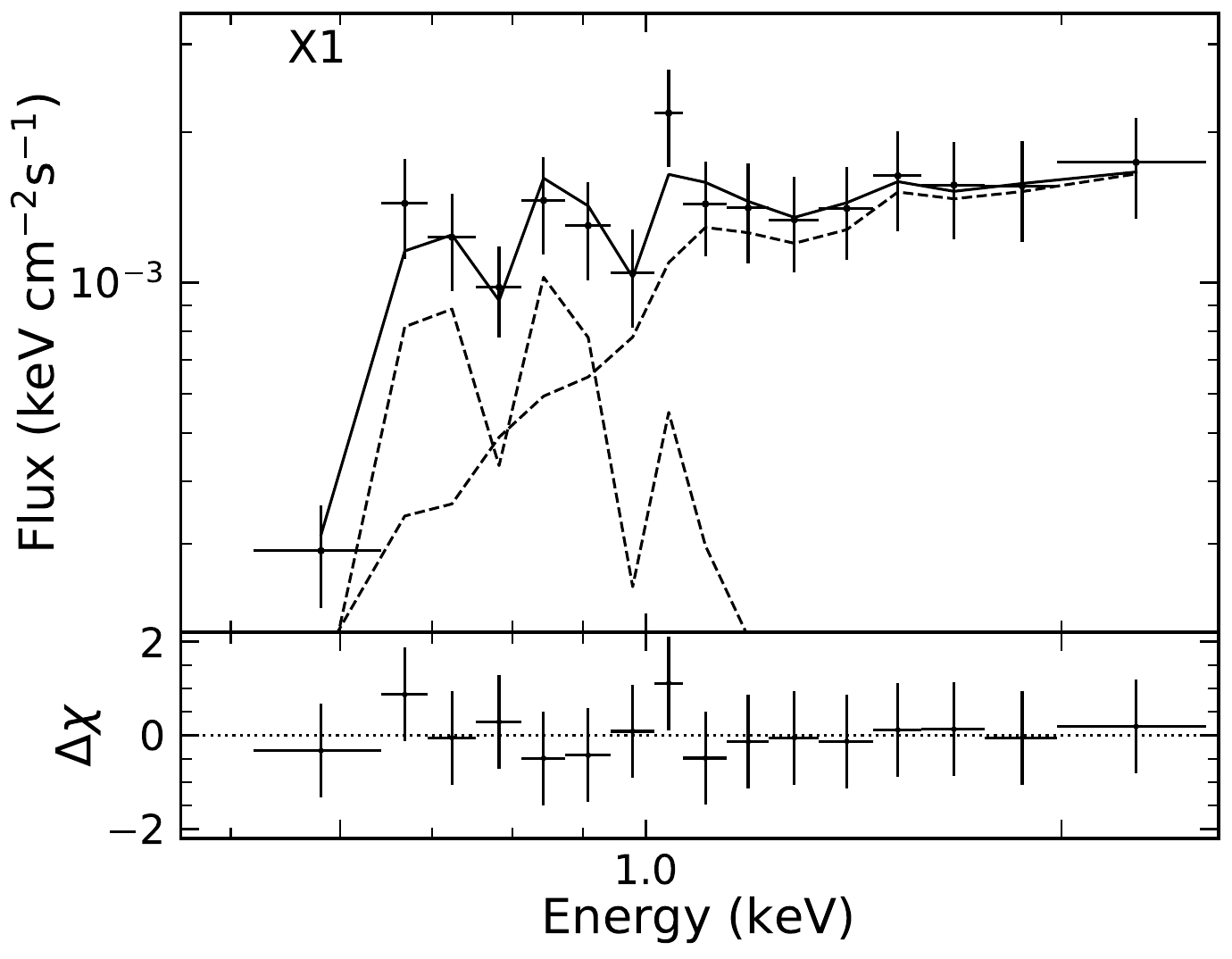}\\
    \vspace*{0.5cm}
    \includegraphics[width=0.65\columnwidth]{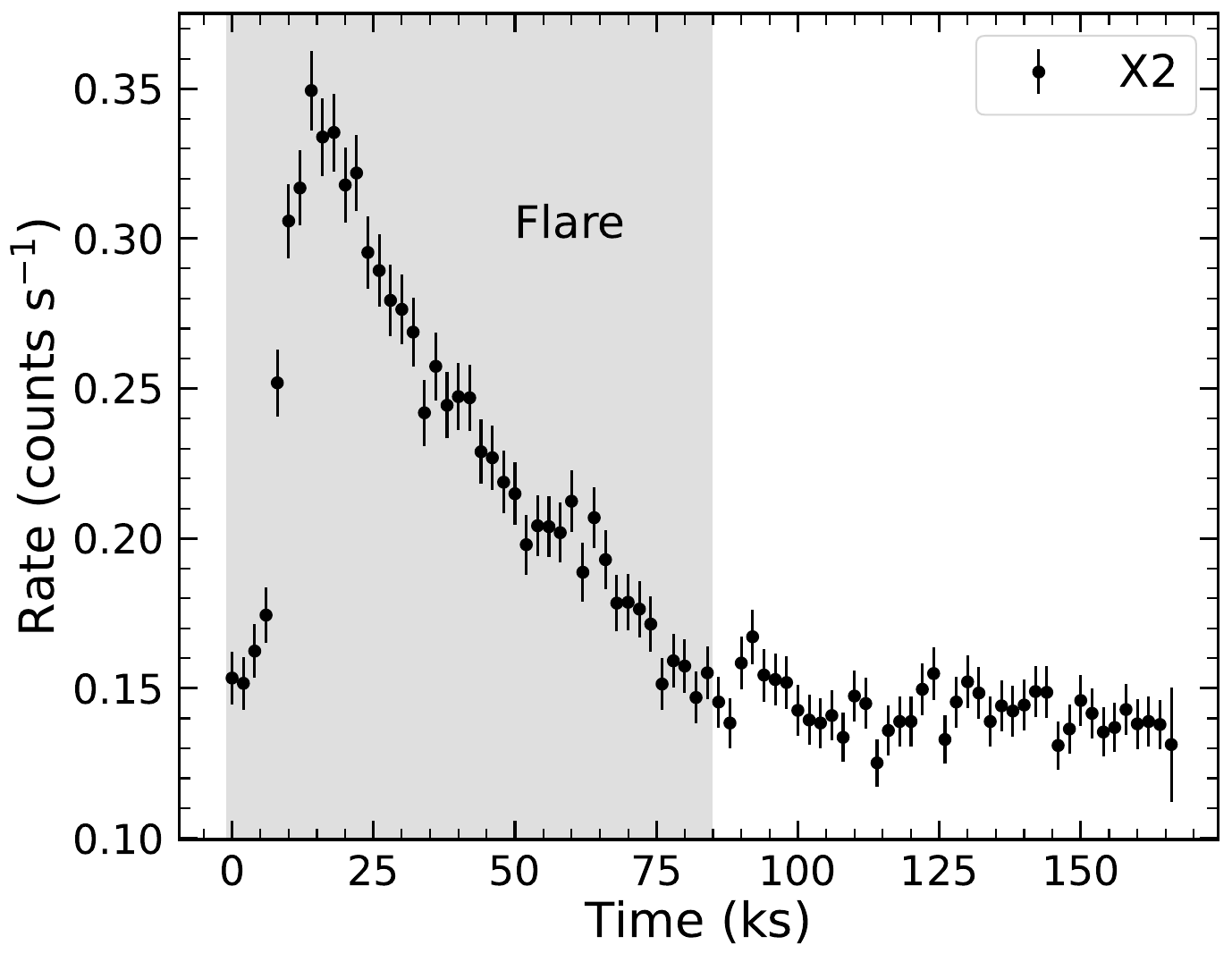}
    \includegraphics[width=0.65\columnwidth]{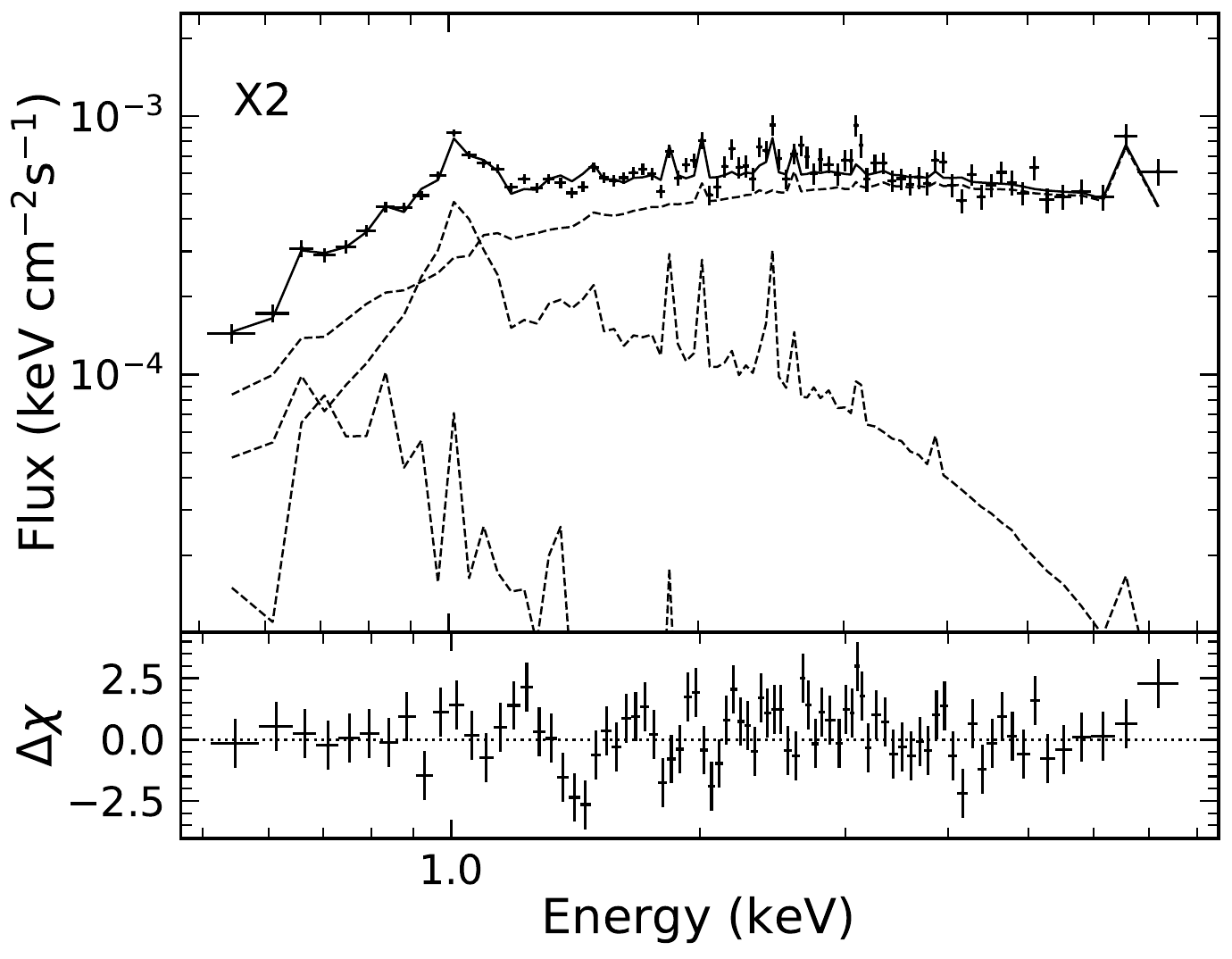}\\
    \vspace*{0.5cm}
    \includegraphics[width=0.66\columnwidth]{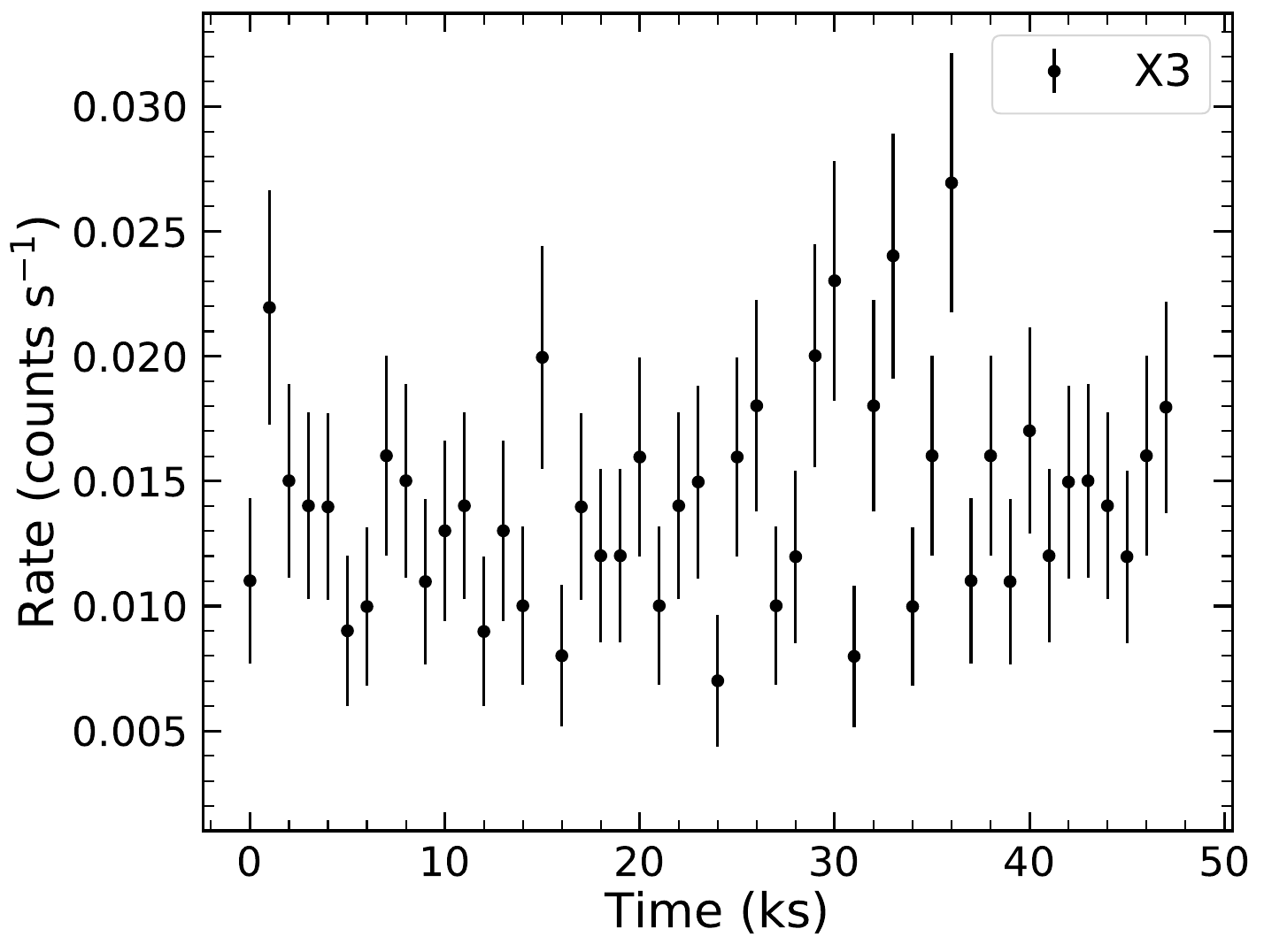}
    \includegraphics[width=0.64\columnwidth]{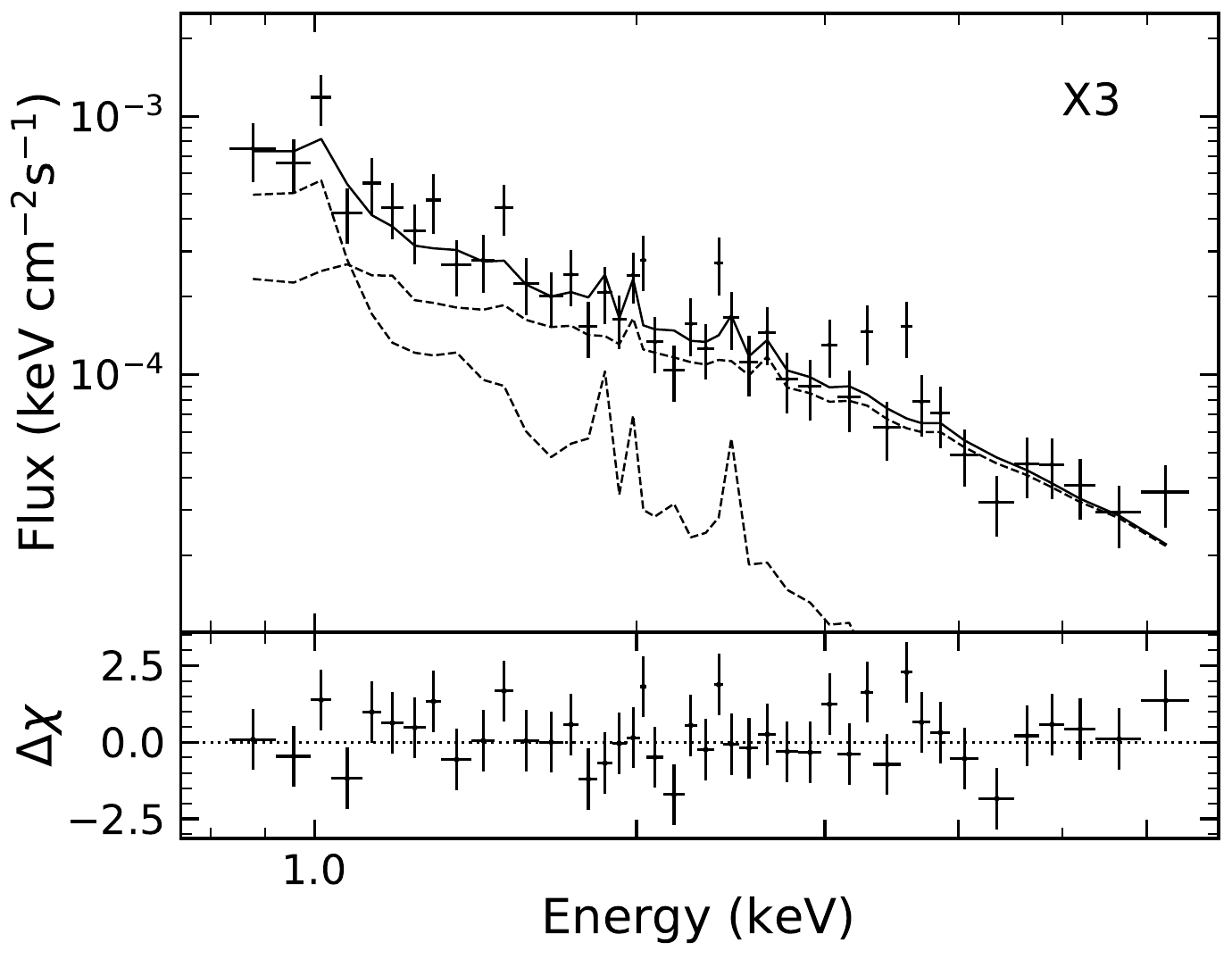}\\
    \vspace*{0.5cm}
    \includegraphics[width=0.66\columnwidth]{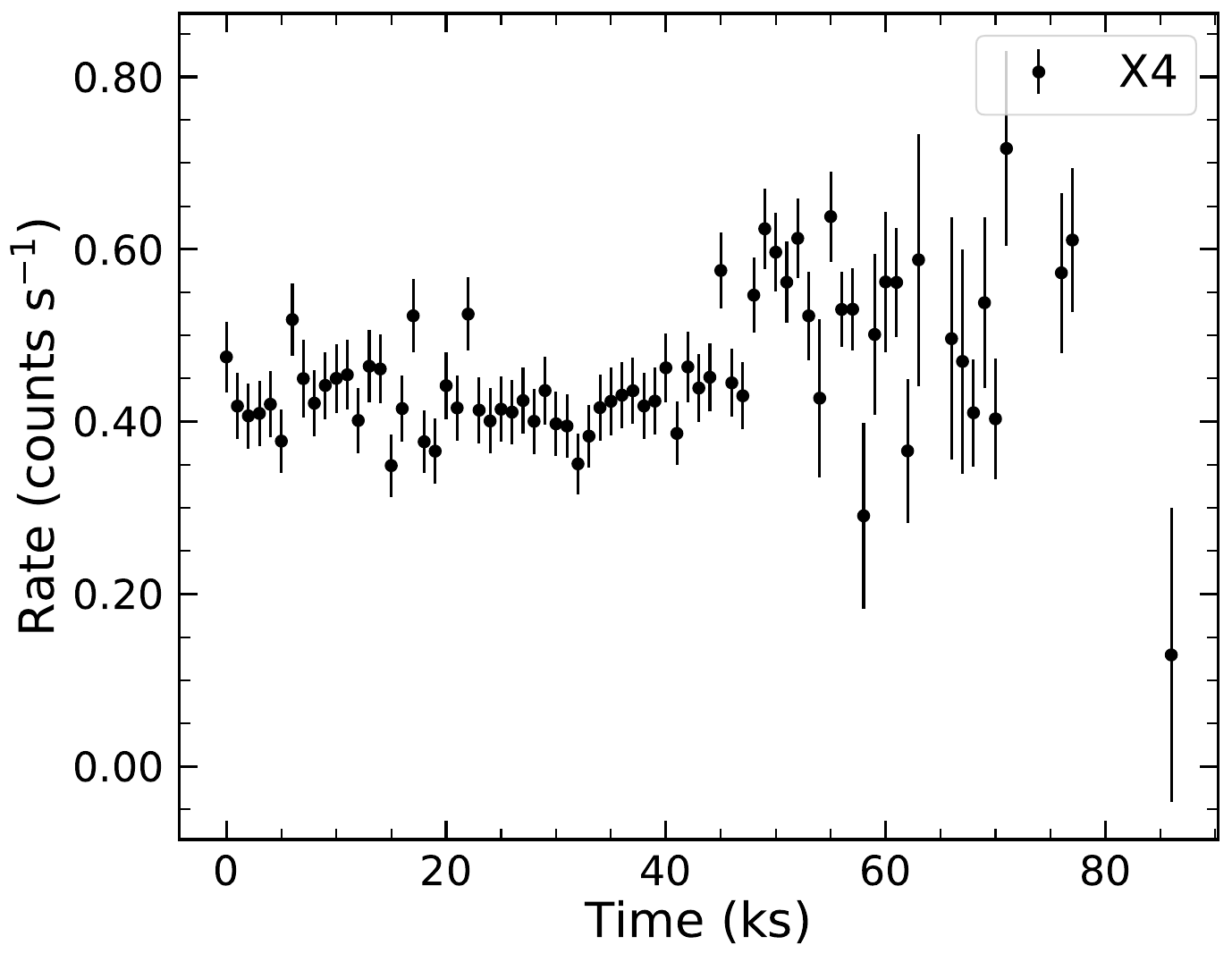}
    \includegraphics[width=0.66\columnwidth]{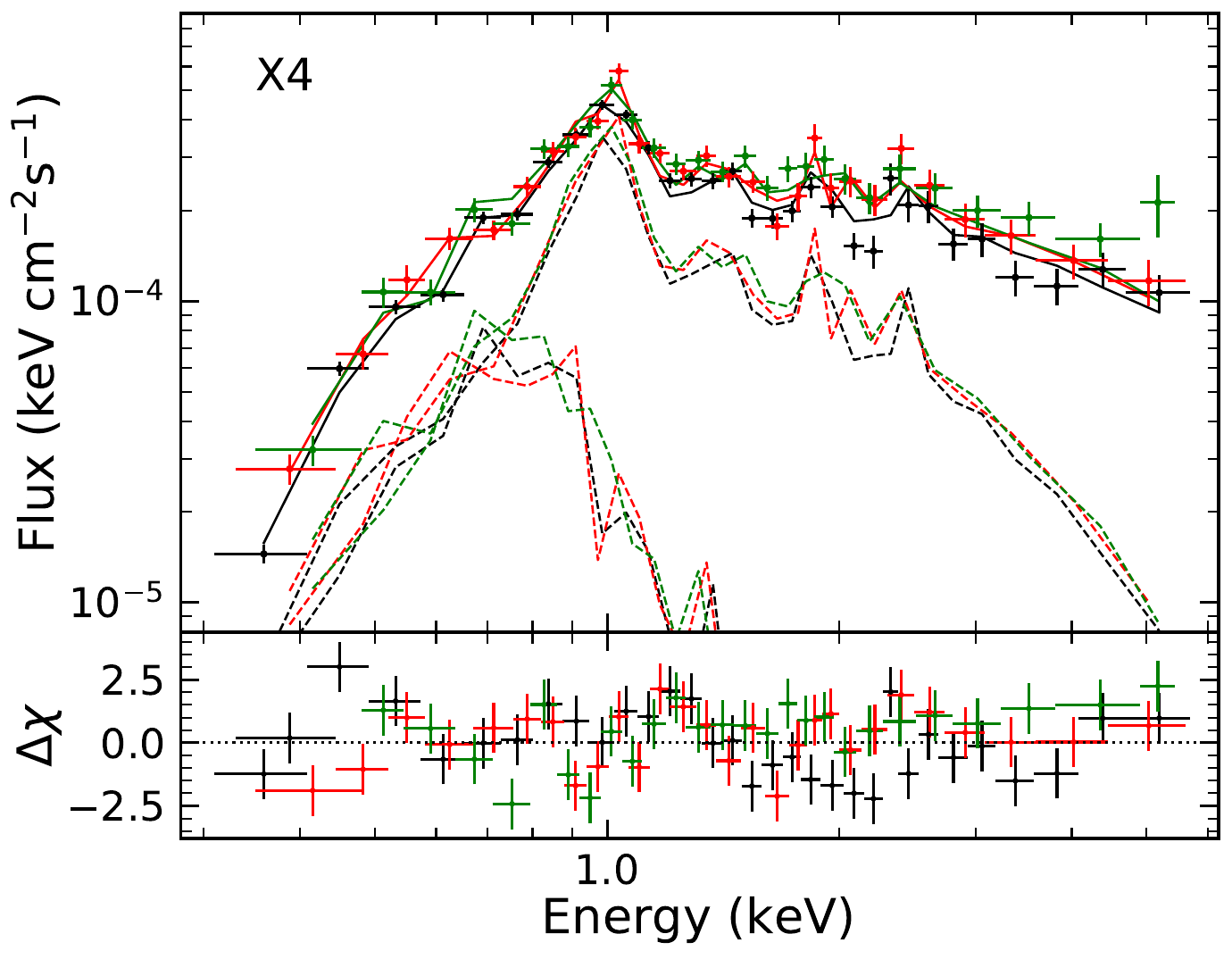}
    \caption{X-ray (0.2--5~keV) light curves (background-subtracted) and spectra for the eROSITA (X1), Chandra (X2 \& X3), and \textit{XMM-Newton} (X4) epochs of observations. The shaded region in the X2 light curve indicates the flaring interval observed in the \textit{Chandra} observation for which we extracted the spectrum. For the X4 epoch, the EPIC/PN light curve is shown. The spectra were fitted with multi-temperature plasma models. In the upper panels of the spectral fitting plots, the data and best-fit models, including individual model components, are displayed. The lower panels show the residuals of the spectral fit. In the XMM-Newton spectra, the black, red, and green points represent the PN, MOS1, and MOS2 data, respectively.}
    \label{fig:xray}
\end{figure*}

\subsubsection{Spectra}
\label{xry_spec}

We analysed the X-ray spectra from observations X1, X2, X3, and X4 using XSPEC version 12.13.0c \citep{1996ASPC..101...17Arnaud} (HEASOFT version 6.31.1). We used the Astrophysical Plasma Emission Code (APEC) for optically thin thermal plasma models \citep{2001ApJ...556L..91Smith} to fit the data. The abundances relative to solar values were set to values from \cite{1998SSRv...85..161Grevesse}. The correction for the Galactic absorption along the line-of-sight was accounted for using the Tuebingen-Boulder ISM absorption model (TBabs). We also checked other abundance libraries and phabs model to account for the Galactic absorption. We found that the $N_H$ value, along with other parameters and flux, are within error bars. Since there is no evidence of circumstellar material in the source, we did not apply any model for intrinsic absorption.

We used the 0.2-5~keV band for eROSITA, 0.3-8~keV for \textit{Chandra} and 0.3-6~keV for \textit{XMM-Newton}, considering the instrument calibration and background dominance. The X-ray spectra for observations X2 and X3 were extracted using the CIAO script {\it specextract} and grouped to a minimum of 25 and 15 counts/bin, respectively. $\chi^2$ statistic was applied as the fit statistic. The uncertainties on the model parameters are at a 90\% confidence level.

We applied time-resolved spectroscopy to the \textit{Chandra} X2 data and extracted the spectra for the flaring state depicted in Fig.\ref{fig:xray}. We required 3 APEC components (3-T model) to fit the \textit{XMM-Newton} (X4) and \textit{Chandra} (X2) spectra, whereas the lower quality spectra of eROSITA (X1) and Chandra (X3) were fitted well with a 2-T model. The absorbing column density ($N_{\rm H}$) was set as a free parameter for all the observations. The best-fit values of N$_H$ fall within the range derived using the extinction law of \cite{1996Ap&SS.236..285Ryter}, i.e., $N_{\rm H}\,[{\rm cm}^{-2}$]\,=\,$A_v$[mag]\,$\times$\,2.0\,$\times$\,10$^{21}$, and were consistent among the observations. The cross-calibration differences between EPIC PN, MOS1 and MOS2 spectra were accounted for using {\sc CONSTANT} model in {\sc xspec}.

We calculated the unabsorbed flux for the 0.2--5~keV band using the convolution model \textit{cflux} in XSPEC. The corresponding flux and the distance (d = 394.2 pc) were then used to determine the X-ray luminosity (${\rm L_{0.2-5keV}}$). We also calculated the emission measure (\textit{EM}) of each plasma component from the normalisation (N) of the APEC model ($EM = 10^{14} \times 4 \pi D^2 N$). The details of the best-fit temperatures and the derived parameters are given in Table \ref{tab:xray_fit}. We found that the luminosities and EMs are higher in the eROSITA observation than in other datasets. This indicates that the source was in a flaring state at the eROSITA epoch. The model components indicate the presence of thermal plasma with temperatures around 0.3~keV and 3~keV in the eROSITA spectrum. An additional plasma component with kT around 1~keV was observed in \textit{Chandra} and \textit{XMM-Newton} spectra.

\begin{figure}
    \centering
    \includegraphics[trim=0cm 0.3cm 0cm 0.2cm, clip=True,scale=0.2]{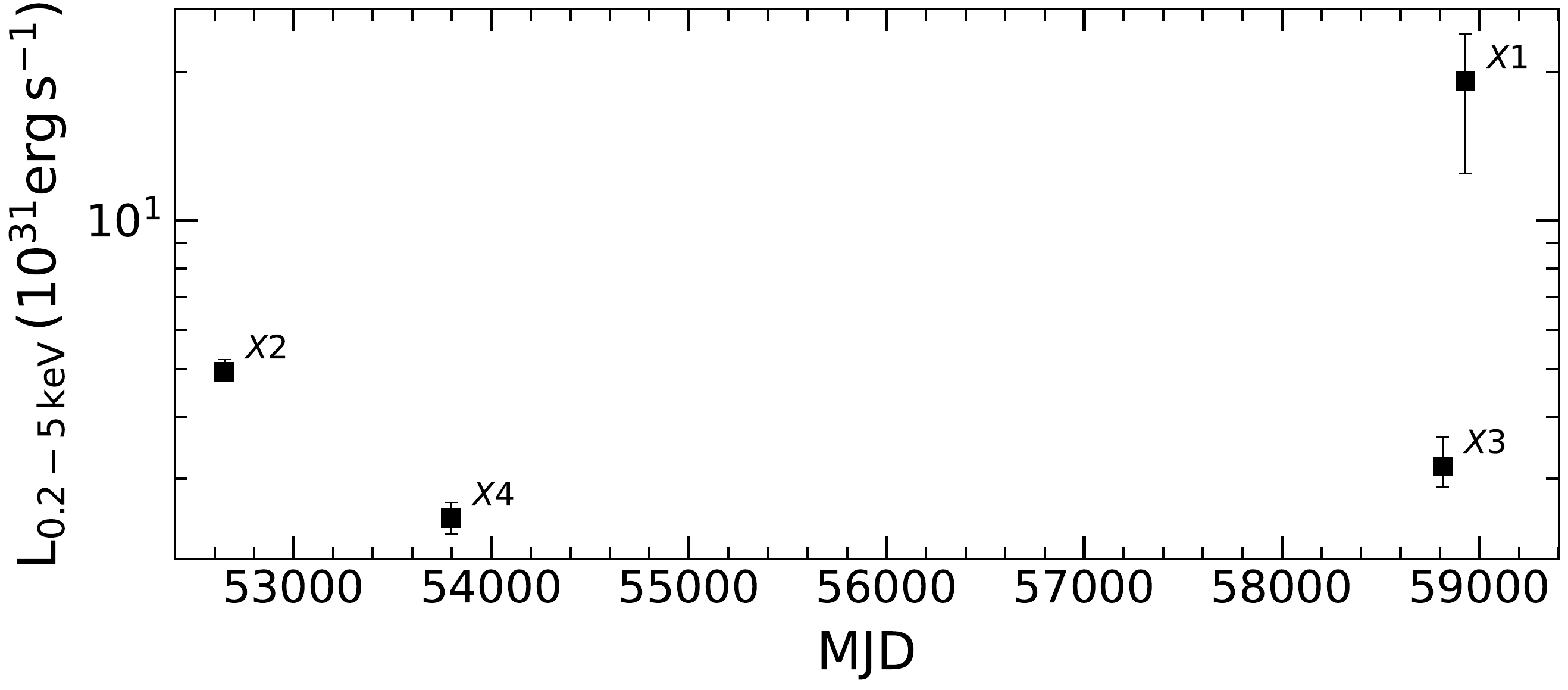}
    \caption{Plot showing the changes in X-ray luminosities obtained with eROSITA (X1), \textit{Chandra} (X2 \& X3) and \textit{XMM-Newton} (X4) epochs.}
    \label{fig:lumin}
\end{figure}

Fig.~\ref{fig:xray} shows the light curves and spectral fitting plots for the X-ray observations with eROSITA, \textit{Chandra} and \textit{XMM-Newton} described above. Fig.~\ref{fig:lumin} displays the X-ray (0.2--5keV) luminosities for these observations.

\subsubsection{Contamination from nearby sources} \label{contamination}

\begin{figure*}[h]
    \centering
    \includegraphics[width=0.25\textwidth]{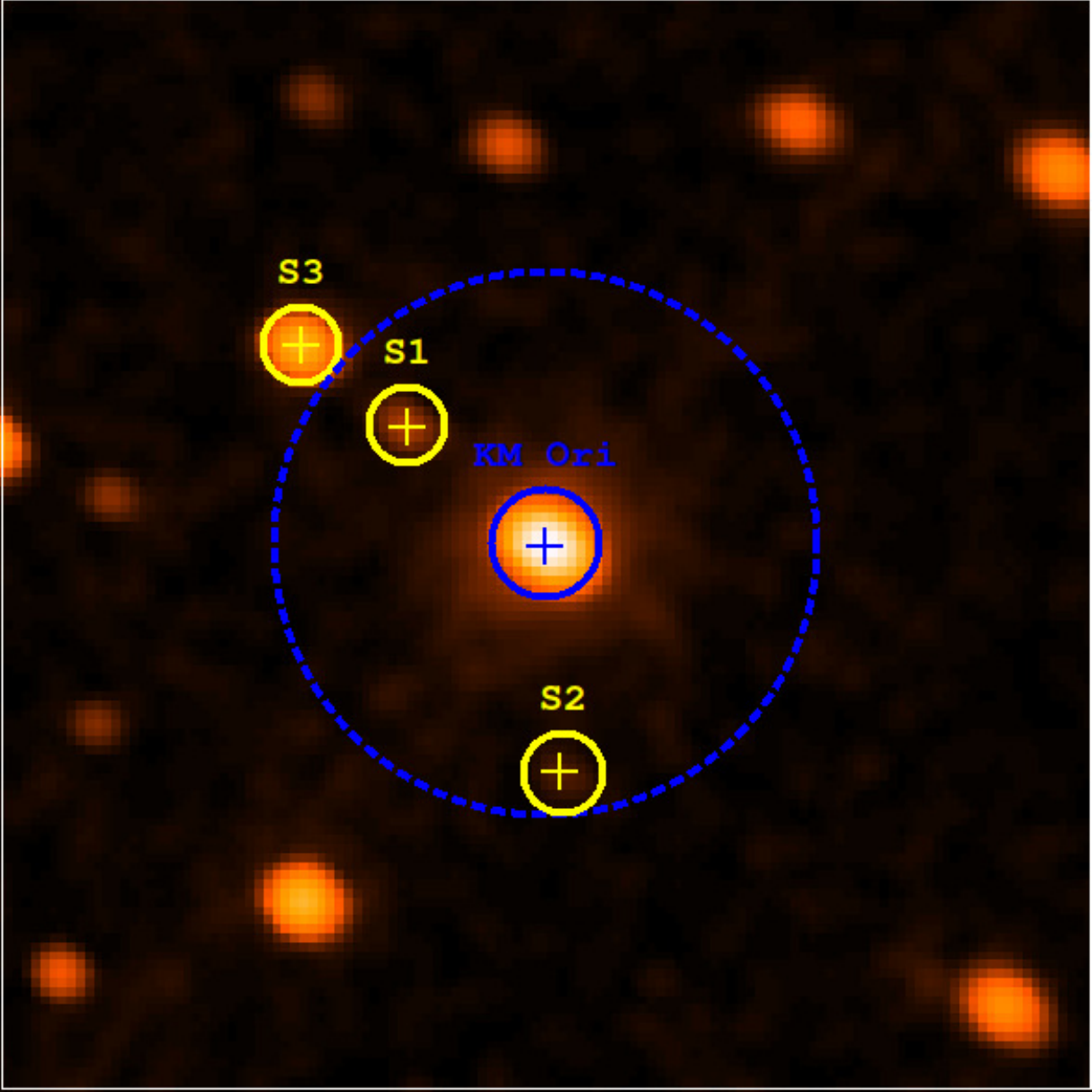}
    \includegraphics[width=0.25\textwidth]{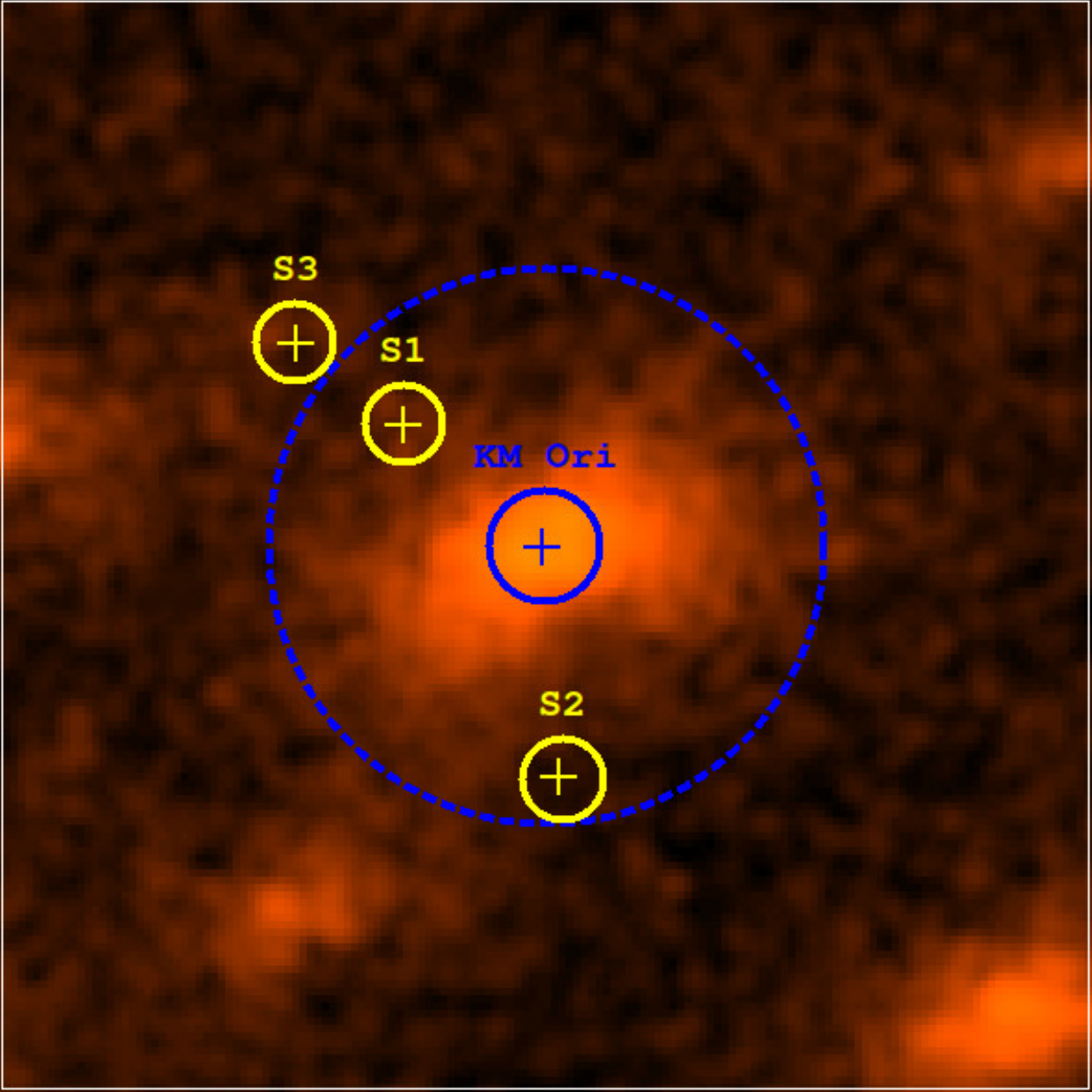}
    \includegraphics[width=0.25\textwidth]{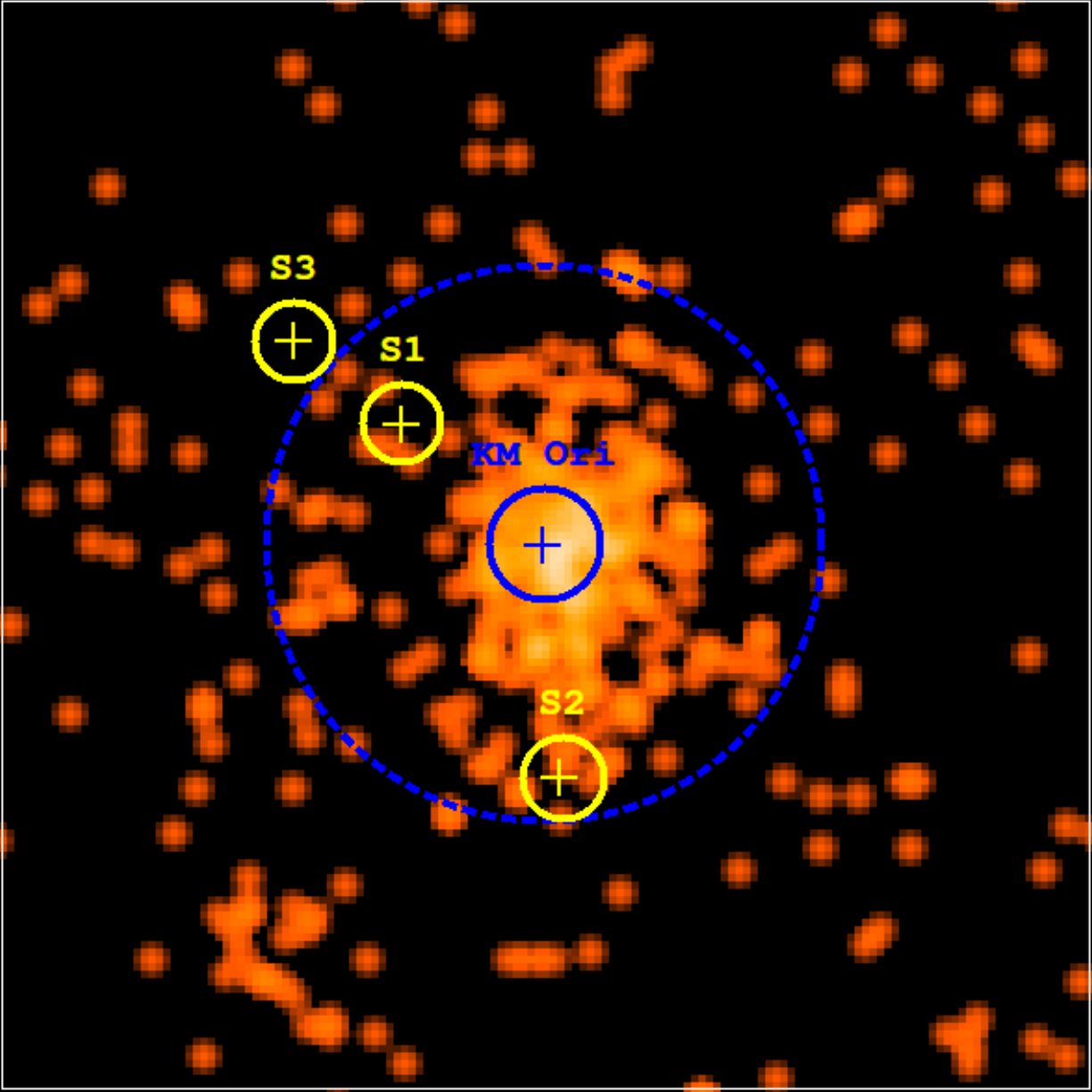}
    \caption{Left - 120x120 arcsec \textit{Chandra} ACIS-I CCD image of KM~Ori and nearby sources (S1, S2 and S3). The image is binned to a pixel size of 0.98 arcsec. Middle - 120x120 arcsec \textit{XMM-Newton} EPIC PN CCD image of KM~Ori and nearby sources (S1, S2 and S3). The image is binned to a pixel size of 0.8 arcsec. Right - 120 arcsec x 120 arcsec \textit{eROSITA} CCD image of KM~Ori observed using all 7 telescope modules. The image is binned to a pixel size of 0.8 arcsec. The blue dashed line circle denotes the extraction region ($\sim$ 30 arcsec) of KM~Ori for the \textit{eROSITA} and \textit{XMM-Newton} data. The blue solid line circle denotes the extraction region ($\sim$~6~arcsec) of KM~Ori for the \textit{Chandra} data. The nearby sources S1, S2 and S3 are marked with yellow crosses. Their coordinates in \textit{XMM-Newton} and \textit{eROSITA} images are fixed at \textit{Chandra} coordinates.}
    \label{fig:CCD_images}
\end{figure*}

The angular resolution of \textit{eROSITA} is $\sim$16~arcsec, and that of EPIC cameras onboard \textit{XMM-Newton} telescope is $\sim$6~arcsec. The ACIS camera onboard \textit{Chandra} has an angular resolution of $\sim$0.5~arcsecs. The distinct resolution of each mission is evident from Fig.~\ref{fig:CCD_images}. Due to the limited angular resolution of \textit{eROSITA} and \textit{XMM-Newton}, the FOV of KM\,Ori appears crowded. In the case of eROSITA, a spurious flag value of 1 (flag{\_}sp{\_}scl=1) is assigned to the KM~Ori data, indicating a crowded region. This crowding raises the possibility of contamination in KM~Ori from nearby sources. Interestingly, three distinct sources (labelled as S1, S2 and S3 in Fig.~\ref{fig:CCD_images}) are seen in the Chandra image. Because of the lower spatial resolution of eROSITA and \textit{XMM-Newton}, it is challenging to separate nearby sources from KM~Ori. Sources S1 and S2 are within the extraction region of the \textit{eROSITA} data, while source S3 is outside this extraction region. Hence, we calculated the percentage of flux due to the sources S1 and S2 that could be contributing to the flux of KM\,Ori. Using the \textit{Chandra} data, we first measured the flux of the sources KM\,Ori, S1 and S2 individually. The sources S1 and S2 are very faint (spectral counts < 40), so the flux cannot be obtained from model fitting. Therefore, we used PIMMS to calculate their flux from the count rate in the energy range of 0.5 $-$ 8 keV. Next, we measured the total flux from the three objects KM\,Ori, S1 and S2 considering a circular extraction region of 30~arcsecs (approximately the size of extraction regions considered for eROSITA and \textit{XMM-Newton}). We find the percentage of flux (\% F$_{cont}$ = \(\frac{F_{KM~Ori}}{F_{S1+S2}}\) $\times$ 100) contributed by sources S1 and S2 to be less than 2.5{\%} for the \textit{Chandra} data. The contribution of flux from the sources S1 and S2 is not significant and does not influence the \textit{XMM-Newton} and eROSITA spectra of KM~Ori.

\begin{table*}[h]
    \tabcolsep 3pt
    \centering
    \caption{Best-fit X-ray spectral parameters for the absorbed thermal plasma model and the luminosity in the 0.2--5~keV band.}
    \begin{tabular}{cccccccccc}
    \hline
    \hline
        OBS & $N_H$ & kT$_1$ & EM$_1$ & kT$_2$ & EM$_2$ & kT$_3$ & EM$_3$ & $L_{\rm 0.2-5 keV}$ & $\chi2/DOF$ \\
          & (10$^{22}$cm$^{-2}$) & (keV) & ($10^{54}$\,cm$^{-3}$) & (keV) & ($10^{54}$\,cm$^{-3}$) & (keV) & (10$^{54}$\,cm$^{-3}$) & ($\rm{10^{31} erg~s^{-1}}$) & \\
        \hline
X1 & $ 0.29 ^{+ 0.31 }_{- 0.18 }$& $ 0.32 ^{+ 0.35 }_{- 0.10 }$& $ 7.92 ^{+ 57.76 }_{- 6.79 }$& -- & -- & $ 2.81 ^{+ 12.62 }_{- 1.16 }$& $ 11.44 ^{+ 4.06 }_{- 2.55 }$& $ 19.34 ^{+ 46.82 }_{- 6.89 }$& 2.96 / 11
\\

X2 & $ 0.17 ^{+ 0.05 }_{- 0.03 }$& $ 0.38 ^{+ 0.23 }_{- 0.09 }$& $ 0.36 ^{+ 0.49 }_{- 0.23 }$& $ 1.24 ^{+ 0.07 }_{- 0.06 }$& $ 1.46 ^{+ 0.34 }_{- 0.25 }$& $ 5.02 ^{+ 0.72 }_{- 0.5 }$& $ 2.85 ^{+ 0.22 }_{- 0.25 }$& $ 5.03 ^{+ 0.46 }_{- 0.25 }$& 281.12 / 239
\\
X3 & $< 0.25$ & -- & -- & $ 0.98 ^{+0.55}_{-0.72} $ & $ 1.05 ^{+1.53  }_{-0.48  }$& $ 2.99 ^{+ 2.52 }_{- 0.66 }$ & $ 1.66 ^{+0.85 }_{-1.24 }$& $ 2.88 ^{+ 1.00 }_{- 0.28 }$& 39.76 / 38
\\

X4 & $ 0.17 \pm 0.02$& $ 0.27 ^{+ 0.05 }_{- 0.03 }$& $ 0.53 ^{+ 0.19 }_{- 0.17 }$& $ 1.08 ^{+ 0.06 }_{- 0.04 }$& $ 1.02 ^{+ 0.17 }_{- 0.12 }$& $ 2.93 ^{+ 0.94 }_{- 0.41 }$& $ 0.84 ^{+ 0.14 }_{- 0.19 }$& $ 2.49 ^{+ 0.19 }_{- 0.17 }$& 295.23 / 246
\\
\hline
\hline
    \end{tabular}
    \label{tab:xray_fit}
\end{table*}

\subsection{Confirming the weak-line nature of KM~Ori}

\begin{figure*}[h]
    \centering
    \includegraphics[width=0.65\columnwidth]{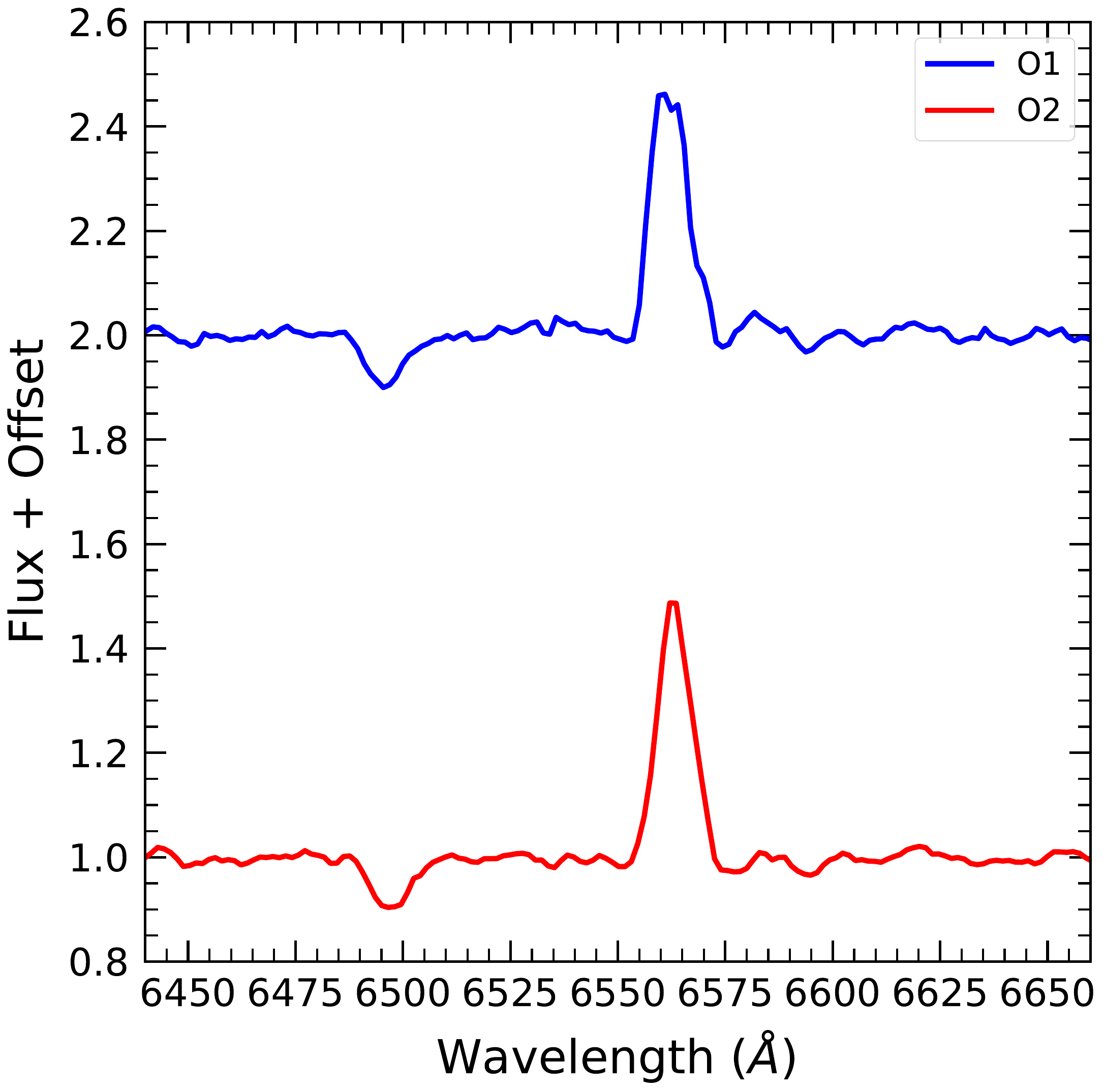}\quad\quad
    \includegraphics[width=0.92\columnwidth]{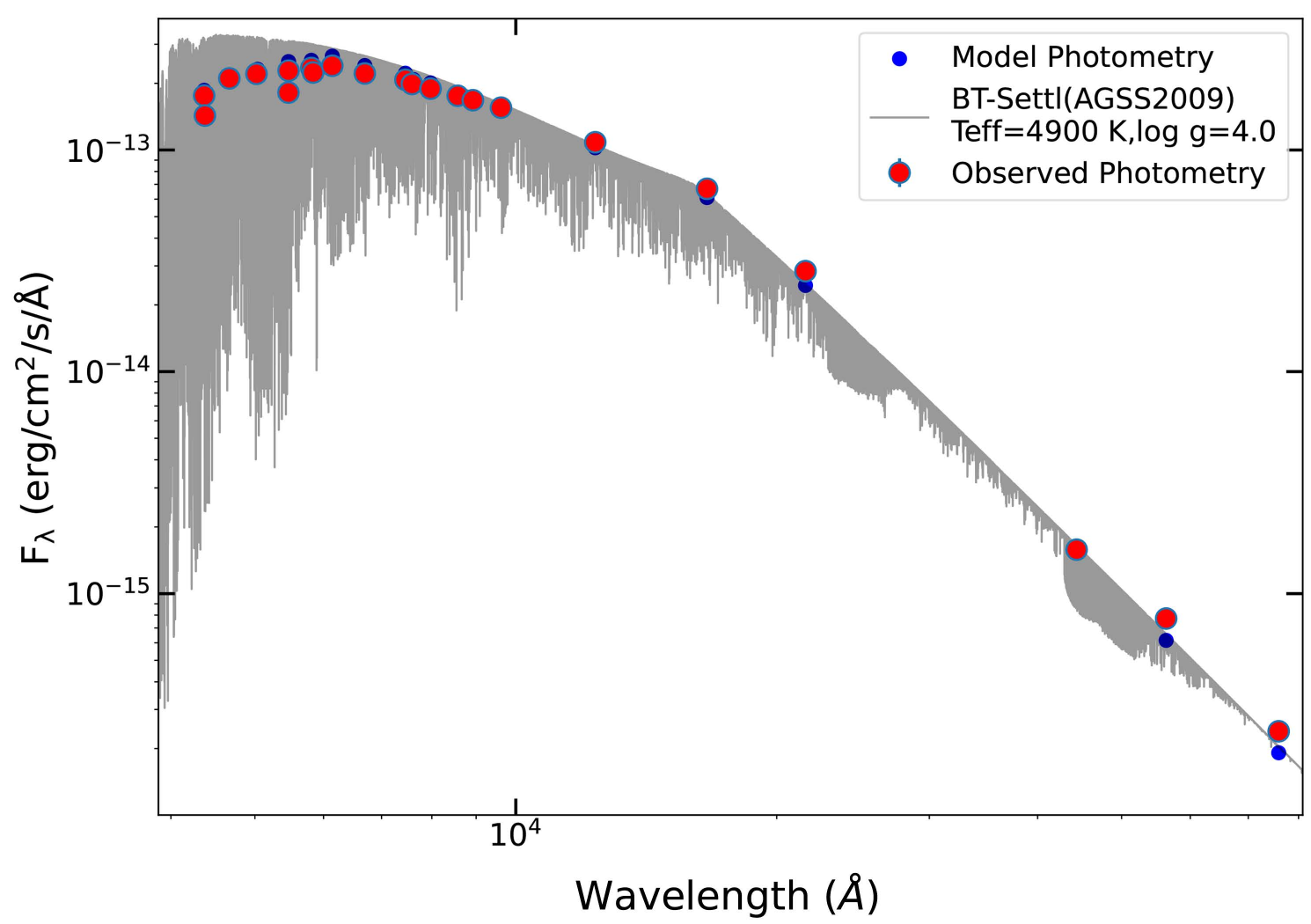}
    \caption{Left: Representation of the H$\alpha$ profiles observed on February 27, 2024 (O1) and March 25, 2024 (O2). Right: SED of KM~Ori fitted with the theoretical BT-Settl model with $A_V$ = 1.05, log(g) = 4, and $T_{\rm eff}$ = 4900~K. The SED shows that KM~Ori does not have any IR excess.}
    \label{fig:hct_spectra_sed}
\end{figure*}

\subsubsection{Optical Spectra}
The primary aim of the optical spectral analysis was to evaluate the H$\alpha$ emission strength of KM~Ori. We observed no significant variation in the H$\alpha$ equivalent width (EW) between the two epochs O1 (-5.1~\AA) and O2 (-4.9~\AA) (see Fig.~\ref{fig:hct_spectra_sed} [Left]). This value falls within the expected range for WTTS, as mentioned by \cite{2005A&A...443..541G}. \cite{1998AJ....115..351M} found a correlation between spectral type and H$\alpha$ EW, concluding that K5 spectral type stars have H$\alpha$ EW equal to or less than -5{\AA}. Based on this analogy, KM~Ori has a spectral type of K5, which supports the spectral type identified in the literature \citep{2019MNRAS.490.3158C}.

\subsubsection{Spectral Energy Distribution}

We constructed the spectral energy distribution (SED) of KM Ori from the available photometry in the Gaia \textit{G}, \textit{$G_{BP}$} and \textit{$G_{RP}$} bands \citep{2023A&A...674A...1G}, the 2MASS \textit{J}, \textit{H} and \textit{K$_s$} bands \citep{2006AJ....131.1163S}, the Spitzer \textit{IRAC} channels \textit{Ch2, Ch3, Ch4} \citep{2004ApJS..154...10F} and Gaia synthetic photometry in the Johnson-Kron-Cousins \textit{BVRI}, the SDSS \textit{griz}, the PANSTARRS \textit{y}, and the HST-ACS-WFC \textit{F606W and F814W} bands \citep{2023A&A...674A..33G}. SED templates corresponding to the queried photometric bands are derived from the BT-Settl model grid \citep{2003IAUS..211..325A, 2011ASPC..448...91A, 2012RSPTA.370.2765A}, spanning from $10^3$~\AA~ to $10^7$~\AA, allowing $T_{eff}$ as a free parameter (2700~K to 70000~K). The best fit was determined by minimising the $\chi^2$ for each instance by using a Python routine \citep{Arun2021MNRAS.507..267A,Shridharan2022A&A...668A.156S,Bhattacharyya2022ApJ...933L..34B}. We fixed the $logg$ parameter in the model grid to a value of 4, as derived from literature for KM Ori \citep{2018AJ....156...84K}. The observed photometry is corrected for extinction using the derived optical extinction ($A_V$ = 1.05 mag) from Green's map \citep{green}. The best-fit value of $T_{eff}$ is 4900~K, which lies in the expected range for a K3-type star. The SED and best-fit theoretical model for KM Ori are displayed in Fig.~\ref{fig:hct_spectra_sed} [Right]. We do not see any excess emission at IR wavelengths, which suggests the lack of dust in the circumstellar environment around the source.

CTTS show variability in the optical band and IR excess in the SED due to the presence of dust in the disk. Our observations of KM Ori have consistently shown an average H$\alpha$ strength of -5~{\AA}. This finding aligns with the typical H$\alpha$ strength observed in WTTS, and no significant variability was detected between the observations conducted on February 27 and March 25, 2024. Furthermore, modelling the SED yields the best fit at $T_{eff} = 4900~K$ and indicates negligible infrared excess up to 5~$\mu$m. This absence of excess emission is consistent with a minimal circumstellar disk, further solidifying the classification of KM Ori as a WTTS.

\section{Discussion and Summary}
\label{rslt_disc}

During our analysis of the APOGEE-2 survey of the Orion Complex with the eROSITA DR1 dataset, we identified the K5-type WTTS KM~Ori exhibiting an exceptionally elevated X-ray flux state compared to other WTTS objects. The X-ray luminosity observed during the eROSITA observation suggests the occurrence of a significant and unforeseen high X-ray flare state in KM~Ori. We generated the long-term X-ray light curve using Chandra observations spanning from 1999 to 2020. Our analysis of the 20-year-long Chandra light curve revealed a high flaring state in KM~Ori in one of the observations obtained during the COUP survey \citep{2005ApJS..160..319G}. This X-ray flare lasted for $\sim 85$~ks with a peak luminosity ($L_{\rm pk}$; 0.5--8 keV) of around $10^{32} {\rm erg~s}^{-1}$ during the 2003 COUP survey \citep{2008ApJ...688..418G}. The eROSITA observation reveals an X-ray luminosity similar to that of the 2003 COUP flaring observation, suggesting that eROSITA captured the source during a flaring episode. Further, we conducted a systematic X-ray analysis of the source utilising eROSITA data and three previous observations from \textit{Chandra} and \textit{XMM-Newton}, as outlined in \S\ref{xry_spec}. We used these three observations to represent the flaring (X2) and quiescent (X3, X4) states, allowing us to make a comparison with the eROSITA flaring observation of the source.

Our analysis revealed a higher X-ray luminosity in the eROSITA observation compared to other epochs. Remarkably, the X-ray (0.2--5~keV) luminosity exceeds $3.9^{+8.2}_{-1.2}$ times more than that of the \textit{Chandra} observation obtained for the time-averaged flaring interval. This clearly indicates that eROSITA caught the object in its most intense X-ray flaring state recorded to date. The ratio of X-ray to bolometric luminosity, log($L_X/L_{bol}$), an indicator of the coronal X-ray activity, comes around -2.4 to -2.9 for different energy bands. The range of values is similar to those reported during the flaring state of T Tauri stars \citep[e.g.,][]{2001ApJ...557..747I, 2005ApJS..160..423W}. X-ray flaring events have also been reported in other WTTS sources \citep[e.g.,][]{1998ApJ...503..894T, 2003A&A...411..517S, 2004A&A...427..263F, 2009ApJ...703..252J, 2009ApJ...697..493B, 2011PASJ...63S.713U, 2013ApJ...771...70G, 2021ApJ...920...22S}. However, coronal X-ray activity higher than that of KM Ori has been reported in only a few WTTS, such as TWA-7 \citep{2011PASJ...63S.713U} and V773 Tau \citep{1998ApJ...503..894T}.

Pre-main sequence stars have been reported to exhibit X-ray superflares/megaflares with peak luminosities in the range log(L$_X$; erg s$^{-1}$) $\sim$ 31--33  and total radiated energies log(E$_X$; erg) $\sim$ 34--38 \citep{2021ApJ...920..154G}. Nevertheless, such flaring events are difficult to detect with normal pointing observations. During the flaring interval of \textit{Chandra} observation of KM~Ori, we observed a time-averaged X-ray (0.5--8~keV) luminosity of approximately ${\rm 5\times10^{31}~erg~s^{-1}}$. The peak X-ray luminosity for the same observation is $L_{\rm pk} {\rm (0.5-8~keV)} \sim 1.5 \times {\rm10^{32}~erg~s^{-1}}$ \citep{2008ApJ...688..418G}. Interestingly, in the eROSITA observation, the source showed a time-averaged X-ray luminosity of $1.6 \times {\rm10^{32}~erg~s^{-1}}$ in the same energy band, indicating a superflare/megaflare state. Additionally, \cite{2021ApJ...916...32G} found that the duration of these flares can range from 3~ks to 154~ks. Though we observe marginal variability in the eROSITA light curve, it is likely that eROSITA has not captured the complete evolution of the flare in KM~Ori, i.e., quiescent, rise and decay phases (see Fig.~\ref{fig:xray}). The high X-ray luminosity (log($L_{\rm 0.5-8~keV}$; erg s$^{-1}$)=32.32) and the time-averaged energy (log($E_{\rm 0.5-8~keV}$; erg)=37.24) at the eROISTA epoch suggest the source to be in a megaflare state that persisted longer than the \textit{Chandra} flaring episode. The eROSITA flare may be classified as a slow-rise top-flat (SRTF) flare \citep{2008ApJ...688..418G}, where variations appear to occur more slowly than in most flares. We cannot entirely rule out the occurrence of repeated short-term flares during the observation, although detecting them is challenging due to the low cadence of eROSITA. However, such intense, repeated flaring events were not observed in previous Chandra data, suggesting that KM Ori may have remained in an SRTF flare state throughout the eROSITA observation ($\sim$86 ks). These flares, characterised by slow rises, long-duration peaks, and/or very long decays, are similar to the slow-rise long-duration flares observed in sources like COUP 1268 and COUP 597, with durations of ~100 ks and ~150 ks, respectively \citep{2008ApJ...688..418G, 2005ApJS..160..469F}. This is consistent with our source KM~Ori, though we do not have a defined flare shape (rise and decay) in the eROSITA light curve. Such long-duration flares are often attributed to X-ray emission from large, hot coronal loops, with sizes ranging from 1.5~R$_{\star}$ to 5.5~R$_{\star}$ \citep{2005ApJS..160..469F}.

Our analysis of the X-ray spectra of KM~Ori at various epochs reveals the presence of multiple thermal plasma components within its corona with temperatures ranging between $\sim$0.3~keV ($\sim$3~MK) and $\sim$5~keV ($\sim$60~MK). The emission measures for all thermal components are notably higher during the eROSITA epoch. According to the solar-flare model \citep{1999ApJ...526L..49S}, the increase in the emission measure could be attributed to heightened evaporation of chromospheric plasma due to the heating of deposited particles \citep{2019MNRAS.483..917W}. Hence, a higher emission measure indicates KM~Ori to be a highly flaring/variable star. 

Given that KM~Ori lacks an accreting disk (weak-line nature) and has no evidence of binarity \citep{1994AJ....108.1906H, 2013ApJS..208...28S, 2019MNRAS.490.3158C}, it suggests that neither of these factors is responsible for generating significant flares in the source. The low-mass PMS stars exhibit pronounced X-ray emission originating from magnetically confined plasma structures in the corona \citep{Lang2012MNRAS.424.1077L}, driven by magnetic fields generated by internal dynamos. Evolutionary tracks for stars of this mass and age suggest KM~Ori be fully convective \citep{1996MNRAS.280.1071L}, indicating that a turbulent dynamo might be operating in this star. The higher occurrence of large flares in WTTS could be due to the more frequent explosive magnetic reconnection, which leads to plasma heating and higher temperatures \citep{2019MNRAS.483..917W}. The elevated levels of magnetic activity in the source may result from the early dissipation of the interacting disks, leading to rapid rotation and strong magnetic dynamos. A similar mechanism has been suggested for another young TTS (age $\sim$1~Myr) V773 Tau (HD 283447), which exhibits strong X-ray emissions during quiescent states (${\rm \sim 10^{31}~erg~s^{-1}}$) and frequent day-long flares peaking at around ${\rm \sim 10^{32}~erg~s^{-1}}$ \citep{1994ApJ...432..373F, 1997ApJ...486..886S, 1998ApJ...503..894T}.

The identification of the most intense X-ray flare in KM~Ori underscores the capability of SRG/eROSITA to detect X-ray flaring occurrences within the WTTS population. Such flares are infrequent and challenging to detect through targeted observations. Investigating larger samples of WTTS from such surveys holds promise for elucidating the fundamental mechanisms governing magnetic activity and topology in young stellar systems. This work is likely the first study to analyse a long-term X-ray light curve of a WTTS. In future work, we explore the long-term X-ray variability of a sample of WTTS, including additional multi-wavelength observations (Anilkumar et al., in prep.). These observations will provide a better understanding of the variability and nature of their X-ray emissions, as well as the physical characteristics of the emitting regions.

\paragraph{Acknowledgement}
This work is based on data from eROSITA, the soft X-ray instrument aboard SRG, a joint Russian-German science mission supported by the Russian Space Agency (Roskosmos), in the interests of the Russian Academy of Sciences represented by its Space Research Institute (IKI), and the Deutsches Zentrum für Luft- und Raumfahrt (DLR). The SRG spacecraft was built by Lavochkin Association (NPOL) and its subcontractors, and is operated by NPOL with support from the Max Planck Institute for Extraterrestrial Physics (MPE). The development and construction of the eROSITA X-ray instrument was led by MPE, with contributions from the Dr. Karl Remeis Observatory Bamberg \& ECAP (FAU Erlangen-Nuernberg), the University of Hamburg Observatory, the Leibniz Institute for Astrophysics Potsdam (AIP), and the Institute for Astronomy and Astrophysics of the University of Tübingen, with the support of DLR and the Max Planck Society. The Argelander Institute for Astronomy of the University of Bonn and the Ludwig Maximilians Universität Munich also participated in the science preparation for eROSITA. BM, SB, and SSK acknowledge the financial support from CHRIST (Deemed to be University, Bangalore) through the SEED money projects (No: SMSS-2335, 11/2023 \& SMSS-2220,12/2022). The authors acknowledge the support provided by the Department of Science and Technology (DST) under the ‘Fund for Improvement of S \& T Infrastructure (FIST)’ program (SR/FST/PS-I/2022/208). BM and SK acknowledge the support provided by the SERB project (CRG/2023/005271). SE thanks Dr T. Saha for the discussions regarding data analysis. VJ and SSK thank the Inter-University Centre for Astronomy and Astrophysics (IUCAA), Pune, India, for the Visiting Associateship.

\paragraph{Data Availability Statement}
    The data underlying this article will be shared on reasonable request.
\bibliographystyle{apalike}
\bibliography{ref}

\begin{thebibliography}{}

\bibitem[{Alcala} et~al., 1997]{1997A&A...319..184A}
{Alcala}, J.~M., {Krautter}, J., {Covino}, E., {Neuhaeuser}, R., {Schmitt}, J.~H.~M.~M., and {Wichmann}, R. (1997).
\newblock {A study of the Chamaeleon star-forming region from the ROSAT All-Sky Survey. II. The pre-main sequence population.}
\newblock {\em \aap}, 319:184--200.

\bibitem[{Allard} et~al., 2003]{2003IAUS..211..325A}
{Allard}, F., {Guillot}, T., {Ludwig}, H.-G., {Hauschildt}, P.~H., {Schweitzer}, A., {Alexander}, D.~R., and {Ferguson}, J.~W. (2003).
\newblock {Model Atmospheres and Spectra: The Role of Dust}.
\newblock In {Mart{\'\i}n}, E., editor, {\em Brown Dwarfs}, volume 211, page 325.

\bibitem[{Allard} et~al., 2011]{2011ASPC..448...91A}
{Allard}, F., {Homeier}, D., and {Freytag}, B. (2011).
\newblock {Model Atmospheres From Very Low Mass Stars to Brown Dwarfs}.
\newblock In {Johns-Krull}, C., {Browning}, M.~K., and {West}, A.~A., editors, {\em 16th Cambridge Workshop on Cool Stars, Stellar Systems, and the Sun}, volume 448 of {\em Astronomical Society of the Pacific Conference Series}, page~91.

\bibitem[{Allard} et~al., 2012]{2012RSPTA.370.2765A}
{Allard}, F., {Homeier}, D., and {Freytag}, B. (2012).
\newblock {Models of very-low-mass stars, brown dwarfs and exoplanets}.
\newblock {\em Philosophical Transactions of the Royal Society of London Series A}, 370(1968):2765--2777.

\bibitem[{Arcodia} et~al., 2024]{2024A&A...684A..64A}
{Arcodia}, R., {Liu}, Z., {Merloni}, A., {Malyali}, A., {Rau}, A., {Chakraborty}, J., {Goodwin}, A., {Buckley}, D., {Brink}, J., {Gromadzki}, M., {Arzoumanian}, Z., {Buchner}, J., {Kara}, E., {Nandra}, K., {Ponti}, G., {Salvato}, M., {Anderson}, G., {Baldini}, P., {Grotova}, I., {Krumpe}, M., {Maitra}, C., {Miller-Jones}, J.~C.~A., and {Ramos-Ceja}, M.~E. (2024).
\newblock {The more the merrier: SRG/eROSITA discovers two further galaxies showing X-ray quasi-periodic eruptions}.
\newblock {\em \aap}, 684:A64.

\bibitem[{Arcodia} et~al., 2021]{2021Natur.592..704A}
{Arcodia}, R., {Merloni}, A., {Nandra}, K., {Buchner}, J., {Salvato}, M., {Pasham}, D., {Remillard}, R., {Comparat}, J., {Lamer}, G., {Ponti}, G., {Malyali}, A., {Wolf}, J., {Arzoumanian}, Z., {Bogensberger}, D., {Buckley}, D.~A.~H., {Gendreau}, K., {Gromadzki}, M., {Kara}, E., {Krumpe}, M., {Markwardt}, C., {Ramos-Ceja}, M.~E., {Rau}, A., {Schramm}, M., and {Schwope}, A. (2021).
\newblock {X-ray quasi-periodic eruptions from two previously quiescent galaxies}.
\newblock {\em \nat}, 592(7856):704--707.

\bibitem[{Arnaud}, 1996]{1996ASPC..101...17Arnaud}
{Arnaud}, K.~A. (1996).
\newblock {XSPEC: The First Ten Years}.
\newblock In {Jacoby}, G.~H. and {Barnes}, J., editors, {\em Astronomical Data Analysis Software and Systems V}, volume 101 of {\em Astronomical Society of the Pacific Conference Series}, page~17.

\bibitem[{Arun} et~al., 2021]{Arun2021MNRAS.507..267A}
{Arun}, R., {Mathew}, B., {Maheswar}, G., {Baug}, T., {Kartha}, S.~S., {Selvakumar}, G., {Manoj}, P., {Shridharan}, B., {Anusha}, R., and {Narang}, M. (2021).
\newblock {Clustering of low-mass stars around Herbig Be star IL Cep - evidence of 'Rocket Effect' using Gaia EDR3 ?}
\newblock {\em \mnras}, 507(1):267--281.

\bibitem[{Baldovin-Saavedra} et~al., 2009]{2009ApJ...697..493B}
{Baldovin-Saavedra}, C., {Audard}, M., {Duch{\^e}ne}, G., {G{\"u}del}, M., {Skinner}, S.~L., {Paerels}, F.~B.~S., {Ghez}, A., and {McCabe}, C. (2009).
\newblock {HDE 245059: A Weak-Lined T Tauri Binary Revealed by Chandra and Keck}.
\newblock {\em \apj}, 697(1):493--505.

\bibitem[{Bhattacharyya} et~al., 2022]{Bhattacharyya2022ApJ...933L..34B}
{Bhattacharyya}, S., {Mathew}, B., {Ezhikode}, S.~H., {Muneer}, S., {Selvakumar}, G., {Maheswer}, G., {Arun}, R., {Anilkumar}, H., {Banerjee}, G., {Pramod}, K.~S., {Kartha}, S.~S., {Paul}, K.~T., and {Velu}, C. (2022).
\newblock {Decoding the X-Ray Flare from MAXI J0709-159 Using Optical Spectroscopy and Multiepoch Photometry}.
\newblock {\em \apjl}, 933(2):L34.

\bibitem[{Bustamante} et~al., 2016]{2016A&A...587A..81B}
{Bustamante}, I., {Mer{\'\i}n}, B., {Bouy}, H., {Manara}, C.~F., {Ribas}, {\'A}., and {Riviere-Marichalar}, P. (2016).
\newblock {X-ray deficiency on strongly accreting T Tauri stars. Comparing Orion with Taurus}.
\newblock {\em \aap}, 587:A81.

\bibitem[{Cruzal{\`e}bes} et~al., 2019]{2019MNRAS.490.3158C}
{Cruzal{\`e}bes}, P., {Petrov}, R.~G., {Robbe-Dubois}, S., {Varga}, J., {Burtscher}, L., {Allouche}, F., {Berio}, P., {Hofmann}, K.~H., {Hron}, J., {Jaffe}, W., {Lagarde}, S., {Lopez}, B., {Matter}, A., {Meilland}, A., {Meisenheimer}, K., {Millour}, F., and {Schertl}, D. (2019).
\newblock {A catalogue of stellar diameters and fluxes for mid-infrared interferometry}.
\newblock {\em \mnras}, 490(3):3158--3176.

\bibitem[{Favata} et~al., 2005]{2005ApJS..160..469F}
{Favata}, F., {Flaccomio}, E., {Reale}, F., {Micela}, G., {Sciortino}, S., {Shang}, H., {Stassun}, K.~G., and {Feigelson}, E.~D. (2005).
\newblock {Bright X-Ray Flares in Orion Young Stars from COUP: Evidence for Star-Disk Magnetic Fields?}
\newblock {\em \apjs}, 160(2):469--502.

\bibitem[{Fazio} et~al., 2004]{2004ApJS..154...10F}
{Fazio}, G.~G., {Hora}, J.~L., {Allen}, L.~E., {Ashby}, M.~L.~N., {Barmby}, P., {Deutsch}, L.~K., {Huang}, J.~S., {Kleiner}, S., {Marengo}, M., {Megeath}, S.~T., {Melnick}, G.~J., {Pahre}, M.~A., {Patten}, B.~M., {Polizotti}, J., {Smith}, H.~A., {Taylor}, R.~S., {Wang}, Z., {Willner}, S.~P., {Hoffmann}, W.~F., {Pipher}, J.~L., {Forrest}, W.~J., {McMurty}, C.~W., {McCreight}, C.~R., {McKelvey}, M.~E., {McMurray}, R.~E., {Koch}, D.~G., {Moseley}, S.~H., {Arendt}, R.~G., {Mentzell}, J.~E., {Marx}, C.~T., {Losch}, P., {Mayman}, P., {Eichhorn}, W., {Krebs}, D., {Jhabvala}, M., {Gezari}, D.~Y., {Fixsen}, D.~J., {Flores}, J., {Shakoorzadeh}, K., {Jungo}, R., {Hakun}, C., {Workman}, L., {Karpati}, G., {Kichak}, R., {Whitley}, R., {Mann}, S., {Tollestrup}, E.~V., {Eisenhardt}, P., {Stern}, D., {Gorjian}, V., {Bhattacharya}, B., {Carey}, S., {Nelson}, B.~O., {Glaccum}, W.~J., {Lacy}, M., {Lowrance}, P.~J., {Laine}, S., {Reach}, W.~T., {Stauffer}, J.~A., {Surace}, J.~A., {Wilson}, G., {Wright}, E.~L., {Hoffman}, A.,
  {Domingo}, G., and {Cohen}, M. (2004).
\newblock {The Infrared Array Camera (IRAC) for the Spitzer Space Telescope}.
\newblock {\em \apjs}, 154(1):10--17.

\bibitem[{Feeney-Johansson} et~al., 2021]{Feeney2021A&A...653A.101F}
{Feeney-Johansson}, A., {Purser}, S.~J.~D., {Ray}, T.~P., {Vidotto}, A.~A., {Eisl{\"o}ffel}, J., {Callingham}, J.~R., {Shimwell}, T.~W., {Vedantham}, H.~K., {Hallinan}, G., and {Tasse}, C. (2021).
\newblock {Detection of coherent low-frequency radio bursts from weak-line T Tauri stars}.
\newblock {\em \aap}, 653:A101.

\bibitem[{Feigelson} et~al., 1994]{1994ApJ...432..373F}
{Feigelson}, E.~D., {Welty}, A.~D., {Imhoff}, C., {Hall}, J.~C., {Etzel}, P.~B., {Phillips}, R.~B., and {Lonsdale}, C.~J. (1994).
\newblock {Multiwavelength Study of the Magnetically Active T Tauri Star HD 283447}.
\newblock {\em \apj}, 432:373.

\bibitem[{Fern{\'a}ndez} et~al., 2004]{2004A&A...427..263F}
{Fern{\'a}ndez}, M., {Stelzer}, B., {Henden}, A., {Grankin}, K., {Gameiro}, J.~F., {Costa}, V.~M., {Guenther}, E., {Amado}, P.~J., and {Rodriguez}, E. (2004).
\newblock {The weak-line T Tauri star V410 Tau. II. A flaring star}.
\newblock {\em \aap}, 427:263--278.

\bibitem[{Flaccomio} et~al., 2003]{2003ApJ...582..382Flaccomio}
{Flaccomio}, E., {Damiani}, F., {Micela}, G., {Sciortino}, S., {Harnden}, F.~R., J., {Murray}, S.~S., and {Wolk}, S.~J. (2003).
\newblock {Chandra X-Ray Observation of the Orion Nebula Cluster. I. Detection, Identification, and Determination of X-Ray Luminosities}.
\newblock {\em \apj}, 582(1):382--397.

\bibitem[{Flaccomio} et~al., 2005]{2005ApJS..160..450F}
{Flaccomio}, E., {Micela}, G., {Sciortino}, S., {Feigelson}, E.~D., {Herbst}, W., {Favata}, F., {Harnden}, F.~R., J., and {Vrtilek}, S.~D. (2005).
\newblock {Rotational Modulation of X-Ray Emission in Orion Nebula Young Stars}.
\newblock {\em \apjs}, 160(2):450--468.

\bibitem[{Freeman} et~al., 2002]{2002ApJS..138..185Freeman}
{Freeman}, P.~E., {Kashyap}, V., {Rosner}, R., and {Lamb}, D.~Q. (2002).
\newblock {A Wavelet-Based Algorithm for the Spatial Analysis of Poisson Data}.
\newblock {\em \apjs}, 138(1):185--218.

\bibitem[Fruscione et~al., 2006]{10.1117/12.671760Fruscione}
Fruscione, A., McDowell, J.~C., Allen, G.~E., Brickhouse, N.~S., Burke, D.~J., Davis, J.~E., Durham, N., Elvis, M., Galle, E.~C., Harris, D.~E., Huenemoerder, D.~P., Houck, J.~C., Ishibashi, B., Karovska, M., Nicastro, F., Noble, M.~S., Nowak, M.~A., Primini, F.~A., Siemiginowska, A., Smith, R.~K., and Wise, M. (2006).
\newblock {CIAO: Chandra's data analysis system}.
\newblock In Silva, D.~R. and Doxsey, R.~E., editors, {\em Observatory Operations: Strategies, Processes, and Systems}, volume 6270, page 62701V. International Society for Optics and Photonics, SPIE.

\bibitem[{Gaia Collaboration} et~al., 2023a]{2023A&A...674A..33G}
{Gaia Collaboration}, {Montegriffo}, P., {Bellazzini}, M., {De Angeli}, F., {Andrae}, R., {Barstow}, M.~A., {Bossini}, D., {Bragaglia}, A., {Burgess}, P.~W., {Cacciari}, C., {Carrasco}, J.~M., {Chornay}, N., {Delchambre}, L., {Evans}, D.~W., {Fouesneau}, M., {Fr{\'e}mat}, Y., {Garabato}, D., {Jordi}, C., {Manteiga}, M., {Massari}, D., {Palaversa}, L., {Pancino}, E., {Riello}, M., {Ruz Mieres}, D., {Sanna}, N., {Santove{\~n}a}, R., {Sordo}, R., {Vallenari}, A., {Walton}, N.~A., {Brown}, A.~G.~A., {Prusti}, T., {de Bruijne}, J.~H.~J., {Arenou}, F., {Babusiaux}, C., {Biermann}, M., {Creevey}, O.~L., {Ducourant}, C., {Eyer}, L., {Guerra}, R., {Hutton}, A., {Klioner}, S.~A., {Lammers}, U.~L., {Lindegren}, L., {Luri}, X., {Mignard}, F., {Panem}, C., {Pourbaix}, D., {Randich}, S., {Sartoretti}, P., {Soubiran}, C., {Tanga}, P., {Bailer-Jones}, C.~A.~L., {Bastian}, U., {Drimmel}, R., {Jansen}, F., {Katz}, D., {Lattanzi}, M.~G., {van Leeuwen}, F., {Bakker}, J., {Casta{\~n}eda}, J., {Fabricius}, C., {Galluccio}, L.,
  {Guerrier}, A., {Heiter}, U., {Masana}, E., {Messineo}, R., {Mowlavi}, N., {Nicolas}, C., {Nienartowicz}, K., {Pailler}, F., {Panuzzo}, P., {Riclet}, F., {Roux}, W., {Seabroke}, G.~M., {Th{\'e}venin}, F., {Gracia-Abril}, G., {Portell}, J., {Teyssier}, D., {Altmann}, M., {Audard}, M., {Bellas-Velidis}, I., {Benson}, K., {Berthier}, J., {Blomme}, R., {Busonero}, D., {Busso}, G., {C{\'a}novas}, H., {Carry}, B., {Cellino}, A., {Cheek}, N., {Clementini}, G., {Damerdji}, Y., {Davidson}, M., {de Teodoro}, P., {Nu{\~n}ez Campos}, M., {Dell'Oro}, A., {Esquej}, P., {Fern{\'a}ndez-Hern{\'a}ndez}, J., {Fraile}, E., {Garc{\'\i}a-Lario}, P., {Gosset}, E., {Haigron}, R., {Halbwachs}, J.~L., {Hambly}, N.~C., {Harrison}, D.~L., {Hern{\'a}ndez}, J., {Hestroffer}, D., {Hodgkin}, S.~T., {Holl}, B., {Jan{\ss}en}, K., {Jevardat de Fombelle}, G., {Jordan}, S., {Krone-Martins}, A., {Lanzafame}, A.~C., {L{\"o}ffler}, W., {Marchal}, O., {Marrese}, P.~M., {Moitinho}, A., {Muinonen}, K., {Osborne}, P., {Pauwels}, T., {Recio-Blanco},
  A., {Reyl{\'e}}, C., {Rimoldini}, L., {Roegiers}, T., {Rybizki}, J., {Sarro}, L.~M., {Siopis}, C., {Smith}, M., {Sozzetti}, A., {Utrilla}, E., {van Leeuwen}, M., {Abbas}, U., {{\'A}brah{\'a}m}, P., {Abreu Aramburu}, A., {Aerts}, C., {Aguado}, J.~J., {Ajaj}, M., {Aldea-Montero}, F., {Altavilla}, G., {{\'A}lvarez}, M.~A., {Alves}, J., {Anderson}, R.~I., {Anglada Varela}, E., {Antoja}, T., {Baines}, D., {Baker}, S.~G., {Balaguer-N{\'u}{\~n}ez}, L., {Balbinot}, E., {Balog}, Z., {Barache}, C., {Barbato}, D., {Barros}, M., {Bartolom{\'e}}, S., {Bassilana}, J.~L., {Bauchet}, N., {Becciani}, U., {Berihuete}, A., {Bernet}, M., {Bertone}, S., {Bianchi}, L., {Binnenfeld}, A., {Blanco-Cuaresma}, S., {Boch}, T., {Bombrun}, A., {Bouquillon}, S., {Bramante}, L., {Breedt}, E., {Bressan}, A., {Brouillet}, N., {Brugaletta}, E., {Bucciarelli}, B., {Burlacu}, A., {Butkevich}, A.~G., {Buzzi}, R., {Caffau}, E., {Cancelliere}, R., {Cantat-Gaudin}, T., {Carballo}, R., {Carlucci}, T., {Carnerero}, M.~I., {Casamiquela}, L.,
  {Castellani}, M., {Castro-Ginard}, A., {Chaoul}, L., {Charlot}, P., {Chemin}, L., {Chiaramida}, V., {Chiavassa}, A., {Comoretto}, G., {Contursi}, G., {Cooper}, W.~J., {Cornez}, T., {Cowell}, S., {Crifo}, F., {Cropper}, M., {Crosta}, M., {Crowley}, C., {Dafonte}, C., {Dapergolas}, A., {David}, P., {de Laverny}, P., {De Luise}, F., {De March}, R., {De Ridder}, J., {de Souza}, R., {de Torres}, A., {del Peloso}, E.~F., {del Pozo}, E., {Delbo}, M., {Delgado}, A., {Delisle}, J.~B., {Demouchy}, C., {Dharmawardena}, T.~E., {Diakite}, S., {Diener}, C., {Distefano}, E., {Dolding}, C., {Enke}, H., {Fabre}, C., {Fabrizio}, M., {Faigler}, S., {Fedorets}, G., {Fernique}, P., {Figueras}, F., {Fournier}, Y., {Fouron}, C., {Fragkoudi}, F., {Gai}, M., {Garcia-Gutierrez}, A., {Garcia-Reinaldos}, M., {Garc{\'\i}a-Torres}, M., {Garofalo}, A., {Gavel}, A., {Gavras}, P., {Gerlach}, E., {Geyer}, R., {Giacobbe}, P., {Gilmore}, G., {Girona}, S., {Giuffrida}, G., {Gomel}, R., {Gomez}, A., {Gonz{\'a}lez-N{\'u}{\~n}ez}, J.,
  {Gonz{\'a}lez-Santamar{\'\i}a}, I., {Gonz{\'a}lez-Vidal}, J.~J., {Granvik}, M., {Guillout}, P., {Guiraud}, J., {Guti{\'e}rrez-S{\'a}nchez}, R., {Guy}, L.~P., {Hatzidimitriou}, D., {Hauser}, M., {Haywood}, M., {Helmer}, A., {Helmi}, A., {Sarmiento}, M.~H., {Hidalgo}, S.~L., {H{\l}adczuk}, N., {Hobbs}, D., {Holland}, G., {Huckle}, H.~E., {Jardine}, K., {Jasniewicz}, G., {Jean-Antoine Piccolo}, A., {Jim{\'e}nez-Arranz}, {\'O}., {Juaristi Campillo}, J., {Julbe}, F., {Karbevska}, L., {Kervella}, P., {Khanna}, S., {Kordopatis}, G., {Korn}, A.~J., {K{\'o}sp{\'a}l}, {\'A}., {Kostrzewa-Rutkowska}, Z., {Kruszy{\'n}ska}, K., {Kun}, M., {Laizeau}, P., {Lambert}, S., {Lanza}, A.~F., {Lasne}, Y., {Le Campion}, J.~F., {Lebreton}, Y., {Lebzelter}, T., {Leccia}, S., {Leclerc}, N., {Lecoeur-Taibi}, I., {Liao}, S., {Licata}, E.~L., {Lindstr{\'o}m}, H.~E.~P., {Lister}, T.~A., {Livanou}, E., {Lobel}, A., {Lorca}, A., {Loup}, C., {Madrero Pardo}, P., {Magdaleno Romeo}, A., {Managau}, S., {Mann}, R.~G., {Marchant}, J.~M.,
  {Marconi}, M., {Marcos}, J., {Marcos Santos}, M.~M.~S., {Mar{\'\i}n Pina}, D., {Marinoni}, S., {Marocco}, F., {Marshall}, D.~J., {Martin Polo}, L., {Mart{\'\i}n-Fleitas}, J.~M., {Marton}, G., {Mary}, N., {Masip}, A., {Mastrobuono-Battisti}, A., {Mazeh}, T., {McMillan}, P.~J., {Messina}, S., {Michalik}, D., {Millar}, N.~R., {Mints}, A., {Molina}, D., {Molinaro}, R., {Moln{\'a}r}, L., {Monari}, G., {Mongui{\'o}}, M., {Montero}, A., {Mor}, R., {Mora}, A., {Morbidelli}, R., {Morel}, T., {Morris}, D., {Muraveva}, T., {Murphy}, C.~P., {Musella}, I., {Nagy}, Z., {Noval}, L., {Oca{\~n}a}, F., {Ogden}, A., {Ordenovic}, C., {Osinde}, J.~O., {Pagani}, C., {Pagano}, I., {Palicio}, P.~A., {Pallas-Quintela}, L., {Panahi}, A., {Payne-Wardenaar}, S., {Pe{\~n}alosa Esteller}, X., {Penttil{\"a}}, A., {Pichon}, B., {Piersimoni}, A.~M., {Pineau}, F.~X., {Plachy}, E., {Plum}, G., {Poggio}, E., {Pr{\v{s}}a}, A., {Pulone}, L., {Racero}, E., {Ragaini}, S., {Rainer}, M., {Raiteri}, C.~M., {Ramos}, P., {Ramos-Lerate}, M., {Re
  Fiorentin}, P., {Regibo}, S., {Richards}, P.~J., {Rios Diaz}, C., {Ripepi}, V., {Riva}, A., {Rix}, H.~W., {Rixon}, G., {Robichon}, N., {Robin}, A.~C., {Robin}, C., {Roelens}, M., {Rogues}, H.~R.~O., {Rohrbasser}, L., {Romero-G{\'o}mez}, M., {Rowell}, N., {Royer}, F., {Rybicki}, K.~A., {Sadowski}, G., {S{\'a}ez N{\'u}{\~n}ez}, A., {Sagrist{\`a} Sell{\'e}s}, A., {Sahlmann}, J., {Salguero}, E., {Samaras}, N., {Sanchez Gimenez}, V., {Sarasso}, M., {Schultheis}, M.~S., {Sciacca}, E., {Segol}, M., {Segovia}, J.~C., {S{\'e}gransan}, D., {Semeux}, D., {Shahaf}, S., {Siddiqui}, H.~I., {Siebert}, A., {Siltala}, L., {Silvelo}, A., {Slezak}, E., {Slezak}, I., {Smart}, R.~L., {Snaith}, O.~N., {Solano}, E., {Solitro}, F., {Souami}, D., {Souchay}, J., {Spagna}, A., {Spina}, L., {Spoto}, F., {Steele}, I.~A., {Steidelm{\"u}ller}, H., {Stephenson}, C.~A., {S{\"u}veges}, M., {Surdej}, J., {Szabados}, L., {Szegedi-Elek}, E., {Taris}, F., {Taylor}, M.~B., {Teixeira}, R., {Tolomei}, L., {Tonello}, N., {Torra}, F., {Torra}, J.,
  {Torralba Elipe}, G., {Trabucchi}, M., {Tsounis}, A.~T., {Turon}, C., {Ulla}, A., {Unger}, N., {Vaillant}, M.~V., {van Dillen}, E., {van Reeven}, W., {Vanel}, O., {Vecchiato}, A., {Viala}, Y., {Vicente}, D., {Voutsinas}, S., {Wevers}, T., {Wyrzykowski}, {\L}., {Yoldas}, A., {Yvard}, P., {Zhao}, H., {Zorec}, J., {Zucker}, S., and {Zwitter}, T. (2023a).
\newblock {Gaia Data Release 3. The Galaxy in your preferred colours: Synthetic photometry from Gaia low-resolution spectra}.
\newblock {\em \aap}, 674:A33.

\bibitem[{Gaia Collaboration} et~al., 2023b]{2023A&A...674A...1G}
{Gaia Collaboration}, {Vallenari}, A., {Brown}, A.~G.~A., {Prusti}, T., {de Bruijne}, J.~H.~J., {Arenou}, F., {Babusiaux}, C., {Biermann}, M., {Creevey}, O.~L., {Ducourant}, C., {Evans}, D.~W., {Eyer}, L., {Guerra}, R., {Hutton}, A., {Jordi}, C., {Klioner}, S.~A., {Lammers}, U.~L., {Lindegren}, L., {Luri}, X., {Mignard}, F., {Panem}, C., {Pourbaix}, D., {Randich}, S., {Sartoretti}, P., {Soubiran}, C., {Tanga}, P., {Walton}, N.~A., {Bailer-Jones}, C.~A.~L., {Bastian}, U., {Drimmel}, R., {Jansen}, F., {Katz}, D., {Lattanzi}, M.~G., {van Leeuwen}, F., {Bakker}, J., {Cacciari}, C., {Casta{\~n}eda}, J., {De Angeli}, F., {Fabricius}, C., {Fouesneau}, M., {Fr{\'e}mat}, Y., {Galluccio}, L., {Guerrier}, A., {Heiter}, U., {Masana}, E., {Messineo}, R., {Mowlavi}, N., {Nicolas}, C., {Nienartowicz}, K., {Pailler}, F., {Panuzzo}, P., {Riclet}, F., {Roux}, W., {Seabroke}, G.~M., {Sordo}, R., {Th{\'e}venin}, F., {Gracia-Abril}, G., {Portell}, J., {Teyssier}, D., {Altmann}, M., {Andrae}, R., {Audard}, M., {Bellas-Velidis},
  I., {Benson}, K., {Berthier}, J., {Blomme}, R., {Burgess}, P.~W., {Busonero}, D., {Busso}, G., {C{\'a}novas}, H., {Carry}, B., {Cellino}, A., {Cheek}, N., {Clementini}, G., {Damerdji}, Y., {Davidson}, M., {de Teodoro}, P., {Nu{\~n}ez Campos}, M., {Delchambre}, L., {Dell'Oro}, A., {Esquej}, P., {Fern{\'a}ndez-Hern{\'a}ndez}, J., {Fraile}, E., {Garabato}, D., {Garc{\'\i}a-Lario}, P., {Gosset}, E., {Haigron}, R., {Halbwachs}, J.~L., {Hambly}, N.~C., {Harrison}, D.~L., {Hern{\'a}ndez}, J., {Hestroffer}, D., {Hodgkin}, S.~T., {Holl}, B., {Jan{\ss}en}, K., {Jevardat de Fombelle}, G., {Jordan}, S., {Krone-Martins}, A., {Lanzafame}, A.~C., {L{\"o}ffler}, W., {Marchal}, O., {Marrese}, P.~M., {Moitinho}, A., {Muinonen}, K., {Osborne}, P., {Pancino}, E., {Pauwels}, T., {Recio-Blanco}, A., {Reyl{\'e}}, C., {Riello}, M., {Rimoldini}, L., {Roegiers}, T., {Rybizki}, J., {Sarro}, L.~M., {Siopis}, C., {Smith}, M., {Sozzetti}, A., {Utrilla}, E., {van Leeuwen}, M., {Abbas}, U., {{\'A}brah{\'a}m}, P., {Abreu Aramburu}, A.,
  {Aerts}, C., {Aguado}, J.~J., {Ajaj}, M., {Aldea-Montero}, F., {Altavilla}, G., {{\'A}lvarez}, M.~A., {Alves}, J., {Anders}, F., {Anderson}, R.~I., {Anglada Varela}, E., {Antoja}, T., {Baines}, D., {Baker}, S.~G., {Balaguer-N{\'u}{\~n}ez}, L., {Balbinot}, E., {Balog}, Z., {Barache}, C., {Barbato}, D., {Barros}, M., {Barstow}, M.~A., {Bartolom{\'e}}, S., {Bassilana}, J.~L., {Bauchet}, N., {Becciani}, U., {Bellazzini}, M., {Berihuete}, A., {Bernet}, M., {Bertone}, S., {Bianchi}, L., {Binnenfeld}, A., {Blanco-Cuaresma}, S., {Blazere}, A., {Boch}, T., {Bombrun}, A., {Bossini}, D., {Bouquillon}, S., {Bragaglia}, A., {Bramante}, L., {Breedt}, E., {Bressan}, A., {Brouillet}, N., {Brugaletta}, E., {Bucciarelli}, B., {Burlacu}, A., {Butkevich}, A.~G., {Buzzi}, R., {Caffau}, E., {Cancelliere}, R., {Cantat-Gaudin}, T., {Carballo}, R., {Carlucci}, T., {Carnerero}, M.~I., {Carrasco}, J.~M., {Casamiquela}, L., {Castellani}, M., {Castro-Ginard}, A., {Chaoul}, L., {Charlot}, P., {Chemin}, L., {Chiaramida}, V., {Chiavassa},
  A., {Chornay}, N., {Comoretto}, G., {Contursi}, G., {Cooper}, W.~J., {Cornez}, T., {Cowell}, S., {Crifo}, F., {Cropper}, M., {Crosta}, M., {Crowley}, C., {Dafonte}, C., {Dapergolas}, A., {David}, M., {David}, P., {de Laverny}, P., {De Luise}, F., {De March}, R., {De Ridder}, J., {de Souza}, R., {de Torres}, A., {del Peloso}, E.~F., {del Pozo}, E., {Delbo}, M., {Delgado}, A., {Delisle}, J.~B., {Demouchy}, C., {Dharmawardena}, T.~E., {Di Matteo}, P., {Diakite}, S., {Diener}, C., {Distefano}, E., {Dolding}, C., {Edvardsson}, B., {Enke}, H., {Fabre}, C., {Fabrizio}, M., {Faigler}, S., {Fedorets}, G., {Fernique}, P., {Fienga}, A., {Figueras}, F., {Fournier}, Y., {Fouron}, C., {Fragkoudi}, F., {Gai}, M., {Garcia-Gutierrez}, A., {Garcia-Reinaldos}, M., {Garc{\'\i}a-Torres}, M., {Garofalo}, A., {Gavel}, A., {Gavras}, P., {Gerlach}, E., {Geyer}, R., {Giacobbe}, P., {Gilmore}, G., {Girona}, S., {Giuffrida}, G., {Gomel}, R., {Gomez}, A., {Gonz{\'a}lez-N{\'u}{\~n}ez}, J., {Gonz{\'a}lez-Santamar{\'\i}a}, I.,
  {Gonz{\'a}lez-Vidal}, J.~J., {Granvik}, M., {Guillout}, P., {Guiraud}, J., {Guti{\'e}rrez-S{\'a}nchez}, R., {Guy}, L.~P., {Hatzidimitriou}, D., {Hauser}, M., {Haywood}, M., {Helmer}, A., {Helmi}, A., {Sarmiento}, M.~H., {Hidalgo}, S.~L., {Hilger}, T., {H{\l}adczuk}, N., {Hobbs}, D., {Holland}, G., {Huckle}, H.~E., {Jardine}, K., {Jasniewicz}, G., {Jean-Antoine Piccolo}, A., {Jim{\'e}nez-Arranz}, {\'O}., {Jorissen}, A., {Juaristi Campillo}, J., {Julbe}, F., {Karbevska}, L., {Kervella}, P., {Khanna}, S., {Kontizas}, M., {Kordopatis}, G., {Korn}, A.~J., {K{\'o}sp{\'a}l}, {\'A}., {Kostrzewa-Rutkowska}, Z., {Kruszy{\'n}ska}, K., {Kun}, M., {Laizeau}, P., {Lambert}, S., {Lanza}, A.~F., {Lasne}, Y., {Le Campion}, J.~F., {Lebreton}, Y., {Lebzelter}, T., {Leccia}, S., {Leclerc}, N., {Lecoeur-Taibi}, I., {Liao}, S., {Licata}, E.~L., {Lindstr{\o}m}, H.~E.~P., {Lister}, T.~A., {Livanou}, E., {Lobel}, A., {Lorca}, A., {Loup}, C., {Madrero Pardo}, P., {Magdaleno Romeo}, A., {Managau}, S., {Mann}, R.~G., {Manteiga}, M.,
  {Marchant}, J.~M., {Marconi}, M., {Marcos}, J., {Marcos Santos}, M.~M.~S., {Mar{\'\i}n Pina}, D., {Marinoni}, S., {Marocco}, F., {Marshall}, D.~J., {Martin Polo}, L., {Mart{\'\i}n-Fleitas}, J.~M., {Marton}, G., {Mary}, N., {Masip}, A., {Massari}, D., {Mastrobuono-Battisti}, A., {Mazeh}, T., {McMillan}, P.~J., {Messina}, S., {Michalik}, D., {Millar}, N.~R., {Mints}, A., {Molina}, D., {Molinaro}, R., {Moln{\'a}r}, L., {Monari}, G., {Mongui{\'o}}, M., {Montegriffo}, P., {Montero}, A., {Mor}, R., {Mora}, A., {Morbidelli}, R., {Morel}, T., {Morris}, D., {Muraveva}, T., {Murphy}, C.~P., {Musella}, I., {Nagy}, Z., {Noval}, L., {Oca{\~n}a}, F., {Ogden}, A., {Ordenovic}, C., {Osinde}, J.~O., {Pagani}, C., {Pagano}, I., {Palaversa}, L., {Palicio}, P.~A., {Pallas-Quintela}, L., {Panahi}, A., {Payne-Wardenaar}, S., {Pe{\~n}alosa Esteller}, X., {Penttil{\"a}}, A., {Pichon}, B., {Piersimoni}, A.~M., {Pineau}, F.~X., {Plachy}, E., {Plum}, G., {Poggio}, E., {Pr{\v{s}}a}, A., {Pulone}, L., {Racero}, E., {Ragaini}, S.,
  {Rainer}, M., {Raiteri}, C.~M., {Rambaux}, N., {Ramos}, P., {Ramos-Lerate}, M., {Re Fiorentin}, P., {Regibo}, S., {Richards}, P.~J., {Rios Diaz}, C., {Ripepi}, V., {Riva}, A., {Rix}, H.~W., {Rixon}, G., {Robichon}, N., {Robin}, A.~C., {Robin}, C., {Roelens}, M., {Rogues}, H.~R.~O., {Rohrbasser}, L., {Romero-G{\'o}mez}, M., {Rowell}, N., {Royer}, F., {Ruz Mieres}, D., {Rybicki}, K.~A., {Sadowski}, G., {S{\'a}ez N{\'u}{\~n}ez}, A., {Sagrist{\`a} Sell{\'e}s}, A., {Sahlmann}, J., {Salguero}, E., {Samaras}, N., {Sanchez Gimenez}, V., {Sanna}, N., {Santove{\~n}a}, R., {Sarasso}, M., {Schultheis}, M., {Sciacca}, E., {Segol}, M., {Segovia}, J.~C., {S{\'e}gransan}, D., {Semeux}, D., {Shahaf}, S., {Siddiqui}, H.~I., {Siebert}, A., {Siltala}, L., {Silvelo}, A., {Slezak}, E., {Slezak}, I., {Smart}, R.~L., {Snaith}, O.~N., {Solano}, E., {Solitro}, F., {Souami}, D., {Souchay}, J., {Spagna}, A., {Spina}, L., {Spoto}, F., {Steele}, I.~A., {Steidelm{\"u}ller}, H., {Stephenson}, C.~A., {S{\"u}veges}, M., {Surdej}, J.,
  {Szabados}, L., {Szegedi-Elek}, E., {Taris}, F., {Taylor}, M.~B., {Teixeira}, R., {Tolomei}, L., {Tonello}, N., {Torra}, F., {Torra}, J., {Torralba Elipe}, G., {Trabucchi}, M., {Tsounis}, A.~T., {Turon}, C., {Ulla}, A., {Unger}, N., {Vaillant}, M.~V., {van Dillen}, E., {van Reeven}, W., {Vanel}, O., {Vecchiato}, A., {Viala}, Y., {Vicente}, D., {Voutsinas}, S., {Weiler}, M., {Wevers}, T., {Wyrzykowski}, {\L}., {Yoldas}, A., {Yvard}, P., {Zhao}, H., {Zorec}, J., {Zucker}, S., and {Zwitter}, T. (2023b).
\newblock {Gaia Data Release 3. Summary of the content and survey properties}.
\newblock {\em \aap}, 674:A1.

\bibitem[{Getman} and {Feigelson}, 2021]{2021ApJ...916...32G}
{Getman}, K.~V. and {Feigelson}, E.~D. (2021).
\newblock {X-Ray Superflares from Pre-main-sequence Stars: Flare Energetics and Frequency}.
\newblock {\em \apj}, 916(1):32.

\bibitem[{Getman} et~al., 2008]{2008ApJ...688..418G}
{Getman}, K.~V., {Feigelson}, E.~D., {Broos}, P.~S., {Micela}, G., and {Garmire}, G.~P. (2008).
\newblock {X-Ray Flares in Orion Young Stars. I. Flare Characteristics}.
\newblock {\em \apj}, 688(1):418--436.

\bibitem[{Getman} et~al., 2021]{2021ApJ...920..154G}
{Getman}, K.~V., {Feigelson}, E.~D., and {Garmire}, G.~P. (2021).
\newblock {X-Ray Superflares from Pre-main-sequence Stars: Flare Modeling}.
\newblock {\em \apj}, 920(2):154.

\bibitem[{Getman} et~al., 2005]{2005ApJS..160..319G}
{Getman}, K.~V., {Flaccomio}, E., {Broos}, P.~S., {Grosso}, N., {Tsujimoto}, M., {Townsley}, L., {Garmire}, G.~P., {Kastner}, J., {Li}, J., {Harnden}, F.~R., J., {Wolk}, S., {Murray}, S.~S., {Lada}, C.~J., {Muench}, A.~A., {McCaughrean}, M.~J., {Meeus}, G., {Damiani}, F., {Micela}, G., {Sciortino}, S., {Bally}, J., {Hillenbrand}, L.~A., {Herbst}, W., {Preibisch}, T., and {Feigelson}, E.~D. (2005).
\newblock {Chandra Orion Ultradeep Project: Observations and Source Lists}.
\newblock {\em \apjs}, 160(2):319--352.

\bibitem[{Ghez} et~al., 1993]{1993AJ....106.2005G}
{Ghez}, A.~M., {Neugebauer}, G., and {Matthews}, K. (1993).
\newblock {The Multiplicity of T Tauri Stars in the Star Forming Regions Taurus-Auriga and Ophiuchus-Scorpius: A 2.2 Micron Speckle Imaging Survey}.
\newblock {\em \aj}, 106:2005.

\bibitem[{Gras-Vel{\'a}zquez} and {Ray}, 2005]{2005A&A...443..541G}
{Gras-Vel{\'a}zquez}, {\`A}. and {Ray}, T.~P. (2005).
\newblock {Weak-line T Tauri stars: circumstellar disks and companions. I. Spectral energy distributions and infrared excesses}.
\newblock {\em \aap}, 443(2):541--556.

\bibitem[Green, 2019]{green}
Green, G. (2019).
\newblock {3D dust map from Green et al. (2019)}.

\bibitem[{Grevesse} and {Sauval}, 1998]{1998SSRv...85..161Grevesse}
{Grevesse}, N. and {Sauval}, A.~J. (1998).
\newblock {Standard Solar Composition}.
\newblock {\em \ssr}, 85:161--174.

\bibitem[{G{\"u}nther} et~al., 2013]{2013ApJ...771...70G}
{G{\"u}nther}, H.~M., {Wolter}, U., {Robrade}, J., and {Wolk}, S.~J. (2013).
\newblock {MN Lup: X-Rays from a Weakly Accreting T Tauri Star}.
\newblock {\em \apj}, 771(1):70.

\bibitem[{Herbst} et~al., 1994]{1994AJ....108.1906H}
{Herbst}, W., {Herbst}, D.~K., {Grossman}, E.~J., and {Weinstein}, D. (1994).
\newblock {Catalogue of UBVRI Photometry of T Tauri Stars and Analysis of the Causes of Their Variability}.
\newblock {\em \aj}, 108:1906.

\bibitem[{Imanishi} et~al., 2001]{2001ApJ...557..747I}
{Imanishi}, K., {Koyama}, K., and {Tsuboi}, Y. (2001).
\newblock {Chandra Observation of the {\ensuremath{\rho}} Ophiuchi Cloud}.
\newblock {\em \apj}, 557(2):747--760.

\bibitem[{Jensen} et~al., 2009]{2009ApJ...703..252J}
{Jensen}, E. L.~N., {Cohen}, D.~H., and {Gagn{\'e}}, M. (2009).
\newblock {No Transition Disk? Infrared Excess, PAH, H$_{2}$, and X-Rays from the Weak-Lined T Tauri Star DoAr 21}.
\newblock {\em \apj}, 703(1):252--269.

\bibitem[{Kounkel} et~al., 2018]{2018AJ....156...84K}
{Kounkel}, M., {Covey}, K., {Su{\'a}rez}, G., {Rom{\'a}n-Z{\'u}{\~n}iga}, C., {Hernandez}, J., {Stassun}, K., {Jaehnig}, K.~O., {Feigelson}, E.~D., {Pe{\~n}a Ram{\'\i}rez}, K., {Roman-Lopes}, A., {Da Rio}, N., {Stringfellow}, G.~S., {Kim}, J.~S., {Borissova}, J., {Fern{\'a}ndez-Trincado}, J.~G., {Burgasser}, A., {Garc{\'\i}a-Hern{\'a}ndez}, D.~A., {Zamora}, O., {Pan}, K., and {Nitschelm}, C. (2018).
\newblock {The APOGEE-2 Survey of the Orion Star-forming Complex. II. Six-dimensional Structure}.
\newblock {\em \aj}, 156(3):84.

\bibitem[{Lang} et~al., 2012]{Lang2012MNRAS.424.1077L}
{Lang}, P., {Jardine}, M., {Donati}, J.-F., {Morin}, J., and {Vidotto}, A. (2012).
\newblock {Coronal structure of low-mass stars}.
\newblock {\em \mnras}, 424(2):1077--1087.

\bibitem[{Lavail} et~al., 2017]{2017A&A...608A..77L}
{Lavail}, A., {Kochukhov}, O., {Hussain}, G.~A.~J., {Alecian}, E., {Herczeg}, G.~J., and {Johns-Krull}, C. (2017).
\newblock {Magnetic fields of intermediate mass T Tauri stars}.
\newblock {\em \aap}, 608:A77.

\bibitem[{Lawson} et~al., 1996]{1996MNRAS.280.1071L}
{Lawson}, W.~A., {Feigelson}, E.~D., and {Huenemoerder}, D.~P. (1996).
\newblock {An improved HR diagram for Chamaeleon I pre-main-sequence stars.}
\newblock {\em \mnras}, 280(4):1071--1088.

\bibitem[{Mart{\'\i}n}, 1998]{1998AJ....115..351M}
{Mart{\'\i}n}, E.~L. (1998).
\newblock {Weak and Post-T Tauri Stars around B-Type Members of the Scorpius-Centaurus OB Association}.
\newblock {\em \aj}, 115(1):351--357.

\bibitem[{Merloni} et~al., 2024]{2024A&A...682A..34M}
{Merloni}, A., {Lamer}, G., {Liu}, T., {Ramos-Ceja}, M.~E., {Brunner}, H., {Bulbul}, E., {Dennerl}, K., {Doroshenko}, V., {Freyberg}, M.~J., {Friedrich}, S., {Gatuzz}, E., {Georgakakis}, A., {Haberl}, F., {Igo}, Z., {Kreykenbohm}, I., {Liu}, A., {Maitra}, C., {Malyali}, A., {Mayer}, M.~G.~F., {Nandra}, K., {Predehl}, P., {Robrade}, J., {Salvato}, M., {Sanders}, J.~S., {Stewart}, I., {Tub{\'\i}n-Arenas}, D., {Weber}, P., {Wilms}, J., {Arcodia}, R., {Artis}, E., {Aschersleben}, J., {Avakyan}, A., {Aydar}, C., {Bahar}, Y.~E., {Balzer}, F., {Becker}, W., {Berger}, K., {Boller}, T., {Bornemann}, W., {Br{\"u}ggen}, M., {Brusa}, M., {Buchner}, J., {Burwitz}, V., {Camilloni}, F., {Clerc}, N., {Comparat}, J., {Coutinho}, D., {Czesla}, S., {Dannhauer}, S.~M., {Dauner}, L., {Dauser}, T., {Dietl}, J., {Dolag}, K., {Dwelly}, T., {Egg}, K., {Ehl}, E., {Freund}, S., {Friedrich}, P., {Gaida}, R., {Garrel}, C., {Ghirardini}, V., {Gokus}, A., {Gr{\"u}nwald}, G., {Grandis}, S., {Grotova}, I., {Gruen}, D., {Gueguen}, A.,
  {H{\"a}mmerich}, S., {Hamaus}, N., {Hasinger}, G., {Haubner}, K., {Homan}, D., {Ider Chitham}, J., {Joseph}, W.~M., {Joyce}, A., {K{\"o}nig}, O., {Kaltenbrunner}, D.~M., {Khokhriakova}, A., {Kink}, W., {Kirsch}, C., {Kluge}, M., {Knies}, J., {Krippendorf}, S., {Krumpe}, M., {Kurpas}, J., {Li}, P., {Liu}, Z., {Locatelli}, N., {Lorenz}, M., {M{\"u}ller}, S., {Magaudda}, E., {Mannes}, C., {McCall}, H., {Meidinger}, N., {Michailidis}, M., {Migkas}, K., {Mu{\~n}oz-Giraldo}, D., {Musiimenta}, B., {Nguyen-Dang}, N.~T., {Ni}, Q., {Olechowska}, A., {Ota}, N., {Pacaud}, F., {Pasini}, T., {Perinati}, E., {Pires}, A.~M., {Pommranz}, C., {Ponti}, G., {Poppenhaeger}, K., {P{\"u}hlhofer}, G., {Rau}, A., {Reh}, M., {Reiprich}, T.~H., {Roster}, W., {Saeedi}, S., {Santangelo}, A., {Sasaki}, M., {Schmitt}, J., {Schneider}, P.~C., {Schrabback}, T., {Schuster}, N., {Schwope}, A., {Seppi}, R., {Serim}, M.~M., {Shreeram}, S., {Sokolova-Lapa}, E., {Starck}, H., {Stelzer}, B., {Stierhof}, J., {Suleimanov}, V., {Tenzer}, C.,
  {Traulsen}, I., {Tr{\"u}mper}, J., {Tsuge}, K., {Urrutia}, T., {Veronica}, A., {Waddell}, S.~G.~H., {Willer}, R., {Wolf}, J., {Yeung}, M.~C.~H., {Zainab}, A., {Zangrandi}, F., {Zhang}, X., {Zhang}, Y., and {Zheng}, X. (2024).
\newblock {The SRG/eROSITA all-sky survey. First X-ray catalogues and data release of the western Galactic hemisphere}.
\newblock {\em \aap}, 682:A34.

\bibitem[{Petrov}, 2021]{2021AcAT....2a...1P}
{Petrov}, P. (2021).
\newblock {Classical T Tauri stars: accretion, wind, dust}.
\newblock {\em Acta Astrophysica Taurica}, 2(1):1--8.

\bibitem[{Petrov}, 2003]{2003Ap.....46..506P}
{Petrov}, P.~P. (2003).
\newblock {T Tauri Stars}.
\newblock {\em Astrophysics}, 46(4):506--529.

\bibitem[{Predehl} et~al., 2021]{2021A&A...647A...1P}
{Predehl}, P., {Andritschke}, R., {Arefiev}, V., {Babyshkin}, V., {Batanov}, O., {Becker}, W., {B{\"o}hringer}, H., {Bogomolov}, A., {Boller}, T., {Borm}, K., {Bornemann}, W., {Br{\"a}uninger}, H., {Br{\"u}ggen}, M., {Brunner}, H., {Brusa}, M., {Bulbul}, E., {Buntov}, M., {Burwitz}, V., {Burkert}, W., {Clerc}, N., {Churazov}, E., {Coutinho}, D., {Dauser}, T., {Dennerl}, K., {Doroshenko}, V., {Eder}, J., {Emberger}, V., {Eraerds}, T., {Finoguenov}, A., {Freyberg}, M., {Friedrich}, P., {Friedrich}, S., {F{\"u}rmetz}, M., {Georgakakis}, A., {Gilfanov}, M., {Granato}, S., {Grossberger}, C., {Gueguen}, A., {Gureev}, P., {Haberl}, F., {H{\"a}lker}, O., {Hartner}, G., {Hasinger}, G., {Huber}, H., {Ji}, L., {Kienlin}, A.~v., {Kink}, W., {Korotkov}, F., {Kreykenbohm}, I., {Lamer}, G., {Lomakin}, I., {Lapshov}, I., {Liu}, T., {Maitra}, C., {Meidinger}, N., {Menz}, B., {Merloni}, A., {Mernik}, T., {Mican}, B., {Mohr}, J., {M{\"u}ller}, S., {Nandra}, K., {Nazarov}, V., {Pacaud}, F., {Pavlinsky}, M., {Perinati}, E.,
  {Pfeffermann}, E., {Pietschner}, D., {Ramos-Ceja}, M.~E., {Rau}, A., {Reiffers}, J., {Reiprich}, T.~H., {Robrade}, J., {Salvato}, M., {Sanders}, J., {Santangelo}, A., {Sasaki}, M., {Scheuerle}, H., {Schmid}, C., {Schmitt}, J., {Schwope}, A., {Shirshakov}, A., {Steinmetz}, M., {Stewart}, I., {Str{\"u}der}, L., {Sunyaev}, R., {Tenzer}, C., {Tiedemann}, L., {Tr{\"u}mper}, J., {Voron}, V., {Weber}, P., {Wilms}, J., and {Yaroshenko}, V. (2021).
\newblock {The eROSITA X-ray telescope on SRG}.
\newblock {\em \aap}, 647:A1.

\bibitem[{Prisinzano} et~al., 2008]{2008ApJ...677..401P}
{Prisinzano}, L., {Micela}, G., {Flaccomio}, E., {Stauffer}, J.~R., {Megeath}, T., {Rebull}, L., {Robberto}, M., {Smith}, K., {Feigelson}, E.~D., {Grosso}, N., and {Wolk}, S. (2008).
\newblock {X-Ray Properties of Protostars in the Orion Nebula}.
\newblock {\em \apj}, 677(1):401--424.

\bibitem[{Ryter}, 1996]{1996Ap&SS.236..285Ryter}
{Ryter}, C.~E. (1996).
\newblock {Interstellar Extinction from Infrared to X-Rays: an Overview}.
\newblock {\em \apss}, 236(2):285--291.

\bibitem[{Shibata} and {Yokoyama}, 1999]{1999ApJ...526L..49S}
{Shibata}, K. and {Yokoyama}, T. (1999).
\newblock {Origin of the Universal Correlation between the Flare Temperature and the Emission Measure for Solar and Stellar Flares}.
\newblock {\em \apjl}, 526(1):L49--L52.

\bibitem[{Shridharan} et~al., 2022]{Shridharan2022A&A...668A.156S}
{Shridharan}, B., {Mathew}, B., {Bhattacharyya}, S., {Robin}, T., {Arun}, R., {Kartha}, S.~S., {Manoj}, P., {Nidhi}, S., {Maheshwar}, G., {Paul}, K.~T., {Narang}, M., and {Himanshu}, T. (2022).
\newblock {Emission line star catalogues post-Gaia DR3. A validation of Gaia DR3 data using the LAMOST OBA emission catalogue}.
\newblock {\em \aap}, 668:A156.

\bibitem[{Skinner} and {G{\"u}del}, 2021]{2021ApJ...920...22S}
{Skinner}, S.~L. and {G{\"u}del}, M. (2021).
\newblock {Chandra X-Ray Observations of V830 Tau: A T Tauri Star Hosting an Evanescent Planet}.
\newblock {\em \apj}, 920(1):22.

\bibitem[{Skinner} et~al., 1997]{1997ApJ...486..886S}
{Skinner}, S.~L., {Guedel}, M., {Koyama}, K., and {Yamauchi}, S. (1997).
\newblock {ASCA Observations of the Barnard 209 Dark Cloud and an Intense X-Ray Flare on V773 Tauri}.
\newblock {\em \apj}, 486(2):886--902.

\bibitem[{Skrutskie} et~al., 2006]{2006AJ....131.1163S}
{Skrutskie}, M.~F., {Cutri}, R.~M., {Stiening}, R., {Weinberg}, M.~D., {Schneider}, S., {Carpenter}, J.~M., {Beichman}, C., {Capps}, R., {Chester}, T., {Elias}, J., {Huchra}, J., {Liebert}, J., {Lonsdale}, C., {Monet}, D.~G., {Price}, S., {Seitzer}, P., {Jarrett}, T., {Kirkpatrick}, J.~D., {Gizis}, J.~E., {Howard}, E., {Evans}, T., {Fowler}, J., {Fullmer}, L., {Hurt}, R., {Light}, R., {Kopan}, E.~L., {Marsh}, K.~A., {McCallon}, H.~L., {Tam}, R., {Van Dyk}, S., and {Wheelock}, S. (2006).
\newblock {The Two Micron All Sky Survey (2MASS)}.
\newblock {\em \aj}, 131(2):1163--1183.

\bibitem[{Smith} et~al., 2001]{2001ApJ...556L..91Smith}
{Smith}, R.~K., {Brickhouse}, N.~S., {Liedahl}, D.~A., and {Raymond}, J.~C. (2001).
\newblock {Collisional Plasma Models with APEC/APED: Emission-Line Diagnostics of Hydrogen-like and Helium-like Ions}.
\newblock {\em \apjl}, 556(2):L91--L95.

\bibitem[{Sokal} et~al., 2020]{2020ApJ...888..116S}
{Sokal}, K.~R., {Johns-Krull}, C.~M., {Mace}, G.~N., {Nofi}, L., {Prato}, L., {Lee}, J.-J., and {Jaffe}, D.~T. (2020).
\newblock {The Mean Magnetic Field Strength of CI Tau}.
\newblock {\em \apj}, 888(2):116.

\bibitem[{Stelzer} et~al., 2003]{2003A&A...411..517S}
{Stelzer}, B., {Fern{\'a}ndez}, M., {Costa}, V.~M., {Gameiro}, J.~F., {Grankin}, K., {Henden}, A., {Guenther}, E., {Mohanty}, S., {Flaccomio}, E., {Burwitz}, V., {Jayawardhana}, R., {Predehl}, P., and {Durisen}, R.~H. (2003).
\newblock {The weak-line T Tauri star V410 Tau. I. A multi-wavelength study of variability}.
\newblock {\em \aap}, 411:517--531.

\bibitem[{Stelzer} et~al., 2000]{Stelzer2000A&A...356..949S}
{Stelzer}, B., {Neuh{\"a}user}, R., and {Hambaryan}, V. (2000).
\newblock {X-ray flares on zero-age- and pre-main sequence stars in Taurus-Auriga-Perseus}.
\newblock {\em \aap}, 356:949--971.

\bibitem[{Str{\"u}der} et~al., 2001]{2001A&A...365L..18S}
{Str{\"u}der}, L., {Briel}, U., {Dennerl}, K., {Hartmann}, R., {Kendziorra}, E., {Meidinger}, N., {Pfeffermann}, E., {Reppin}, C., {Aschenbach}, B., {Bornemann}, W., {Br{\"a}uninger}, H., {Burkert}, W., {Elender}, M., {Freyberg}, M., {Haberl}, F., {Hartner}, G., {Heuschmann}, F., {Hippmann}, H., {Kastelic}, E., {Kemmer}, S., {Kettenring}, G., {Kink}, W., {Krause}, N., {M{\"u}ller}, S., {Oppitz}, A., {Pietsch}, W., {Popp}, M., {Predehl}, P., {Read}, A., {Stephan}, K.~H., {St{\"o}tter}, D., {Tr{\"u}mper}, J., {Holl}, P., {Kemmer}, J., {Soltau}, H., {St{\"o}tter}, R., {Weber}, U., {Weichert}, U., {von Zanthier}, C., {Carathanassis}, D., {Lutz}, G., {Richter}, R.~H., {Solc}, P., {B{\"o}ttcher}, H., {Kuster}, M., {Staubert}, R., {Abbey}, A., {Holland}, A., {Turner}, M., {Balasini}, M., {Bignami}, G.~F., {La Palombara}, N., {Villa}, G., {Buttler}, W., {Gianini}, F., {Lain{\'e}}, R., {Lumb}, D., and {Dhez}, P. (2001).
\newblock {The European Photon Imaging Camera on XMM-Newton: The pn-CCD camera}.
\newblock {\em \aap}, 365:L18--L26.

\bibitem[{Sunyaev} et~al., 2021]{2021A&A...656A.132S}
{Sunyaev}, R., {Arefiev}, V., {Babyshkin}, V., {Bogomolov}, A., {Borisov}, K., {Buntov}, M., {Brunner}, H., {Burenin}, R., {Churazov}, E., {Coutinho}, D., {Eder}, J., {Eismont}, N., {Freyberg}, M., {Gilfanov}, M., {Gureyev}, P., {Hasinger}, G., {Khabibullin}, I., {Kolmykov}, V., {Komovkin}, S., {Krivonos}, R., {Lapshov}, I., {Levin}, V., {Lomakin}, I., {Lutovinov}, A., {Medvedev}, P., {Merloni}, A., {Mernik}, T., {Mikhailov}, E., {Molodtsov}, V., {Mzhelsky}, P., {M{\"u}ller}, S., {Nandra}, K., {Nazarov}, V., {Pavlinsky}, M., {Poghodin}, A., {Predehl}, P., {Robrade}, J., {Sazonov}, S., {Scheuerle}, H., {Shirshakov}, A., {Tkachenko}, A., and {Voron}, V. (2021).
\newblock {SRG X-ray orbital observatory. Its telescopes and first scientific results}.
\newblock {\em \aap}, 656:A132.

\bibitem[{Szegedi-Elek} et~al., 2013]{2013ApJS..208...28S}
{Szegedi-Elek}, E., {Kun}, M., {Reipurth}, B., {P{\'a}l}, A., {Bal{\'a}zs}, L.~G., and {Willman}, M. (2013).
\newblock {A New H{\ensuremath{\alpha}} Emission-line Survey in the Orion Nebula Cluster}.
\newblock {\em \apjs}, 208(2):28.

\bibitem[{Tody}, 1986]{1986SPIE..627..733T}
{Tody}, D. (1986).
\newblock {The IRAF Data Reduction and Analysis System}.
\newblock In {Crawford}, D.~L., editor, {\em Instrumentation in astronomy VI}, volume 627 of {\em Society of Photo-Optical Instrumentation Engineers (SPIE) Conference Series}, page 733.

\bibitem[{Tsuboi} et~al., 1998]{1998ApJ...503..894T}
{Tsuboi}, Y., {Koyama}, K., {Murakami}, H., {Hayashi}, M., {Skinner}, S., and {Ueno}, S. (1998).
\newblock {ASCA Detection of a Superhot 100 Million K X-Ray Flare on the Weak-Lined T Tauri Star V773 Tauri}.
\newblock {\em \apj}, 503(2):894--901.

\bibitem[{Uzawa} et~al., 2011]{2011PASJ...63S.713U}
{Uzawa}, A., {Tsuboi}, Y., {Morii}, M., {Yamazaki}, K., {Kawai}, N., {Matsuoka}, M., {Nakahira}, S., {Serino}, M., {Matsumura}, T., {Mihara}, T., {Tomida}, H., {Ueda}, Y., {Sugizaki}, M., {Ueno}, S., {Daikyuji}, A., {Ebisawa}, K., {Eguchi}, S., {Hiroi}, K., {Ishikawa}, M., {Isobe}, N., {Kawasaki}, K., {Kimura}, M., {Kitayama}, H., {Kohama}, M., {Kotani}, T., {Nakagawa}, Y.~E., {Nakajima}, M., {Negoro}, H., {Ozawa}, H., {Shidatsu}, M., {Sootome}, T., {Sugimori}, K., {Suwa}, F., {Tsunemi}, H., {Usui}, R., {Yamamoto}, T., {Yamaoka}, K., and {Yoshida}, A. (2011).
\newblock {A Large X-Ray Flare from a Single Weak-Lined T Tauri Star TWA-7 Detected with MAXI GSC}.
\newblock {\em \pasj}, 63:S713--S716.

\bibitem[{Walter} et~al., 1988]{1988AJ.....96..297W}
{Walter}, F.~M., {Brown}, A., {Mathieu}, R.~D., {Myers}, P.~C., and {Vrba}, F.~J. (1988).
\newblock {X-Ray Sources in Regions of Star Formation. III. Naked T Tauri Stars Associated with the Taurus-Auriga Complex}.
\newblock {\em \aj}, 96:297.

\bibitem[{Waterfall} et~al., 2019]{2019MNRAS.483..917W}
{Waterfall}, C.~O.~G., {Browning}, P.~K., {Fuller}, G.~A., and {Gordovskyy}, M. (2019).
\newblock {Modelling the radio and X-ray emission from T-Tauri flares}.
\newblock {\em \mnras}, 483(1):917--930.

\bibitem[{Wolk} et~al., 2005]{2005ApJS..160..423W}
{Wolk}, S.~J., {Harnden}, F.~R., J., {Flaccomio}, E., {Micela}, G., {Favata}, F., {Shang}, H., and {Feigelson}, E.~D. (2005).
\newblock {Stellar Activity on the Young Suns of Orion: COUP Observations of K5-7 Pre-Main-Sequence Stars}.
\newblock {\em \apjs}, 160(2):423--449.

\end{thebibliography}
\begin{appendix}
\renewcommand{\thetable}{\Alph{section}\arabic{table}}

\end{appendix}
\end{document}